\documentclass{elsart.new}
\usepackage{natbib}
\usepackage{epsfig}
\begin{document}
\begin{frontmatter}
\title{Particle acceleration in ultra-relativistic oblique shock waves} 
\author[IMP,MPIfR]{A. Meli\thanksref{Mailaddr}} and
\author[IMP]{J.J. Quenby}

\address[IMP]{Astrophysics Group, Blackett Laboratory, Imperial College
of Science, Technology and Medicine, Prince Consort Rd. SW7 2BW,
London, UK.}
\address[MPIfR]{Visiting Max Planck Institut fuer Radioastronomie, Auf dem Huegel 69 53121, 
Bonn, Germany.}
\thanks[Mailaddr]{Corresponding author: a.meli@ic.ac.uk}

\begin{abstract}
We perform Monte Carlo simulations of diffusive shock acceleration at highly 
relativistic oblique shock waves. High upstream flow Lorentz gamma factors ($\Gamma$)
are used, which are relevant to models of ultra relativistic particle shock acceleration in 
Active Galactic Nuclei (AGN) central engines and relativistic 
jets and Gamma Ray Burst (GRB) fireballs. We investigate numerically the  
acceleration properties in the relativistic and ultra relativistic flow regime 
($\Gamma \sim 10-10^{3}$), such as angular distribution, acceleration time
constant, particle energy gain versus number of crossings and spectral 
shapes. We perform calculations for sub-luminal and super-luminal
shocks. For the first case, the dependence on whether or not the scattering is pitch 
angle diffusion or large angle scattering is studied. The large angle
model exhibits a distinctive structure in the basic power-law spectrum which is not nearly so obvious
for small angle scattering. However, both models yield significant 'speed-up' or
faster acceleration rates when compared with the conventional, non-relativistic expression
for the time constant, or alternatively with the time scale $r_{g}/c$ where $r_{g}$ is Larmor radius.
The $\Gamma^{2}$ energization for the first crossing cycle and the significantly large energy gain for 
subsequent crossings as well as the high 'speed up' factors found, are important in supporting the
Vietri and Waxman work on GRB ultra-high energy cosmic ray, neutrino, and gamma-ray output. Secondly, 
for super-luminal shocks, we calculate the energy gain for a number
of different inclinations and the spectral shapes of the accelerated 
particles are given. In this investigation we consider only large angle scattering, partly because 
of computational time limitations and partly because this model provides the most favourable situation 
for acceleration. We use high gamma flows with Lorentz factors in the range 10-40, which 
are relevant to AGN accretion disks and jet ultra relativistic shock configurations. 
We closely follow the particle's trajectory along the magnetic field lines during shock crossings 
where the equivalent of a guiding centre approximation is inappropriate, constantly  measuring its 
phase space co-ordinates in the fluid frames where {\bf E}=0. We find that a super-luminal 
'shock drift' mechanism is less efficient in accelerating particles to the highest energies observed, 
compared to the first order Fermi acceleration applying in the sub-luminal case, suggesting that the former 
cannot stand as a sole acceleration mechanism for the ultra-high energy cosmic rays observed.
\end{abstract}

\end{frontmatter}


\section{Introduction}

Detection of $10^{6}-10^{20}$eV nuclei within the solar system, implying their presence
through much of the observable universe, has stimulated the development 
of the diffusive shock acceleration model,
whereby particles are repeatedly accelerated in multiple crossings 
of a shock interface, due to collisions with upstream and downstream magnetic scattering 
centres. The pioneering work of the late 70s (Krymsky, 1977; Bell, 1978a,b; Axford et al., 1978; 
Blandford and Ostriker, 1978) established the basic mechanism of particle diffusive acceleration 
in non-relativistic flows.  However, since then, extension of the model to oblique shocks, 
relativistic plasma flows and the incorporation of  non-linear effects, have rendered the emergence 
of a consensus physical picture more difficult. At oblique shock fronts the motion of the particles is 
more complicated, compared to a  parallel shock configuration, as they may either be transmitted or 
reflected by the shock surface. Apart from the diffusive acceleration mechanism, an oblique shock 
could also accelerate particles in the \textit{absence} of scattering (Armstrong and Decker, 1979), 
and this mechanism is called 'shock drift' acceleration. Provided that the motion of the particles 
is diffusive in the oblique shock configurations it has been shown, for an analytical non-relativistic 
flow approach, that the same result as in parallel shocks, connects the power-law index of
accelerated particles with the compression ratio and the acceleration rate with the diffusion coefficients
in the shock normal direction (e.g. Axford et al., 1978; Bell, 1978a; Drury, 1983).\\
Also, following energy gain in the shock frame on shock crossing via the drift approach, is equivalent 
to following the gain via Lorentz transformations from the shock frame into an {\bf E}=0 frame 
(defined by de Hoffmann and Teller, 1950) at the shock, and 
then back into the shock frame. On the other hand, the theory of oblique shocks considers 
features that are not involved in the parallel shock case, in particular 
the ability of the shock to reflect particles meeting a weaker to stronger field transition and the 
dependence of the acceleration time on the field-shock normal angle (e.g. Jokipii, 1982).
Oblique shocks can be classified into sub-luminal and super-luminal. The difference
between these two categories is that, in the sub-luminal case, it is possible to find a
relativistic transformation to the frame of reference (de Hoffmann-Teller frame), in which
the shock front is stationary and the electric field is zero ({\bf E}=0) in both 
upstream and downstream regions.  
On the other hand, super-luminal shock fronts do not admit a transformation to such a
frame, as they correspond to shock fronts in which the point of intersection of the 
front with a magnetic field line moves with a speed greater than $c$.
Kirk and Heavens (1989) employed a numerical 
solution to the transport equation (in the test particle limit) and
found spectral flattening at relativistic speeds in inclined shocks as well as an increased
anisotropy both in fluid frames and the {\bf E}=0 frame. These results were confirmed by Monte Carlo 
simulations (Ostrowski, 1991 and Lieu et al., 1994). 
The former author investigated the gyromotion across the shock to check the region of applicability 
of magnetic moment conservation in the {\bf E}=0 frame at the shock while, the later authors simply 
relied on this assumption. Baring et al. (1994) noticed 'bumps' in the accelerated spectra 
just upstream of the shock in the non-relativistic oblique case. The reason a Monte Carlo approach 
became favoured, lies in the large anisotropies and gradients which develop around the shock at relativistic 
flow speeds. These render direct solution of the transport equation 
difficult, because the Boltzmann equation must be applied to a fine-grained bundle of particle 
trajectories in phase space and the diffusion approximation, employing a simple spatial diffusion tensor 
is inadequate (e.g. Axford, 1981).\\
All the work mentioned so far is sub-luminal, but Bednarz and Ostrowski (1998) used a small-angle 
scattering model to include  super-luminal situations with field-shock normal angles, 
$\psi \leq \pi/2 $ and flow $\Gamma \leq 243$ and also a simulation of varying amounts of 
cross-field diffusion. These authors found spectral indices which could be very large at intermediate 
$\Gamma$ values with small $K_{\perp}$, but at high $\Gamma$, all results appeared to tend towards a 
differential exponent of -2.2. \\ 
Earlier discussion of an extreme example of a super-luminal case, the pulsar wind shock acceleration 
scheme for the termination shock of an isolated pulsar magnetosphere where the $\Gamma \sim 10^{4}$ 
polar wind meets the slowly expanding nebula  envelope, has emphasized the apparent difficulty in 
achieving multiple shock crossings of accelerated particles. 
This can be seen if we remember that for super-luminal shocks, a transformation to a frame with 
{\bf B} parallel to the shock plane is possible (Hudson, 1965) 
so in the absence of scattering close to the shock, particles are swept at an angle to 
the shock normal into the shock, shock drift and then exit 
downstream with no return possible. Begelman and Kirk (1990) note that only shock compression 
of the upstream population results in the low turbulence case. 
However, Lucek and Bell (1994) find special trajectories if the field is
ordered but not uniform, allowing energy increases by factors of the order of hundreds per
crossing, which coupled with particle motion in the electromagnetic field of the nebula
can render isolated pulsars accelerators of high energy cosmic rays.
Gallant and Arons (1994) provide a rare example of a hybrid model of relativistic particle
acceleration, applied to the Crab Nebula, using a fluid approach for an $e^{\pm}$ pair plasma together
with a kinematic treatment of ions to compute the synchrotron spectrum. \\
Acceleration time under non-relativistic theory clearly depends on shock obliquity
through variation of $K_{n}$, the diffusion coefficient along the shock normal, with $\psi$, where
$K_{n}=K_{||}cos^{2}\psi+K_{\perp}sin^{2}\psi$. $K_{||},K_{\perp}$ are respectively the diffusion 
coefficients parallel and perpendicular to the field. The acceleration time is given by,

\begin{equation}
\tau=\frac{3}{V_{1}-V_{2}}\left(\frac{K_{n,1}}{V_{1}}+\frac{K_{n,2}}{V_{2}}\right)
\end{equation}

Here the suffixes '1' and '2' refer to upstream and downstream, and $V$ is plasma flow velocity. 
Because this equation (1) is clearly an inadequate measure of acceleration time in the highly relativistic
case, we note it may also be written as $\tau$ lying in the range $(4\rightarrow 8/3)(nr_{g}/c)cos^{2}\psi$ 
for a parallel mean free path $\lambda_{\|}=nr_{g}$, a shock frame compression ratio of 4 and a 
downstream reduction in $\lambda_{\|}$ by a factor 4. Hence while it will be convenient to measure any 
'speed-up' of acceleration relative
to equation (1), we may equally regard the time to move over the relativistic Larmor radius as the unit
of acceleration time for comparison purposes. 
Quenby and Lieu (1989) and Ellison et al. (1990) employed Monte Carlo
guiding centre approximation  calculations to show that relativistic parallel shock flows 
with $\Gamma\sim 10$ produced a 'speed-up' of acceleration by factors $\sim3\rightarrow10$ 
when compared with equation (1) for $\tau$. The former authors argued for large angle scattering while 
the latter found essentially the same result with either large or small angle scattering. 
Lieu et al. (1994) showed the speed-up as a function of flow velocity was similar for inclined,  
sub-luminal shocks, provided the results are plotted  against flow seen in the {\bf E}=0  frame.  
Quenby and Drolias (1995) using an inclined shock model which included
the expected shock structure due to cosmic ray pressure back reaction, found some speed-up 
still occurred. Bednarz and Ostrowski (1996) employ a Monte Carlo  
computation scheme to investigate acceleration times for shock speeds up to 0.9$c$ for the inclined case, 
with both  small angle and large angle scattering and allowing for cross 
field motion of guiding centres. Speed-up is noticed especially at 
high shock speed and turbulence although, cross-field diffusion eventually 
reverses the correlation  with turbulence level. Power law spectra only occur in the 
super-luminal case for isotropic scattering. 
Acceleration time scales, shorter than the upstream gyroperiod have been obtained.\\    
In a companion work, Meli and Quenby (2002a), have studied parallel shock acceleration for flow 
$\Gamma \leq 10^{3}$,  for both small and large angle scattering models using the guiding centre Monte 
Carlo approach. They confirm a $\Gamma^{2}$ energy enhancement for the first shock crossing cycle, 
first emphasized by Quenby and Lieu (1989,) and they find significant energy gains subsequently.
Also, the structure in the accelerated particle spectrum at the shock, indicated in previous lower $\Gamma$ 
simulations, became enhanced for large angle scattering, although not in the small angle case,
rendering the idea of a simple power law output spectrum for photon production uncertain and 
they confirm the existence of the speed-up effect.    
It is the purpose of this work to extend the parallel shock studies of Meli and Quenby (2002a)  
to provide answers concerning the acceleration time scales, particle spectral shape at the source, 
angular distribution and energy gain per cycle of the accelerated
particles for different shock inclinations using very high gamma flows. We consider both sub-luminal shocks 
(where an {\bf E}=0 frame can be found) and super-luminal ones (where {\bf E}=0 is impossible) and 
the helical particle trajectories are followed  in the appropriate
frames to determine the characteristics of the shock crossing.
Streaming instabilities, drift waves and Rayleigh-Taylor like instabilities all can increase 
scattering around the shock and contribute to the back reaction of the relativistic gas, but these are not 
enough to destroy the basic relativistic beaming in flows, where $V \rightarrow c$.
Hence a Monte Carlo approach is employed.
The speed-up for a true discontinuity in the test-particle strong scattering limit, is due 
to a $\sim \Gamma ^{2}$ change in energy in one complete shock crossing cycle for shock $\Gamma$,
and this effect is the basic cause of the deviation of the spectral shape from the 
non-relativistic result. Our current interest, 
regarding a real astrophysical scenario, lies in relativistic flows 
in AGN jets where $\Gamma\approx 5-10 $ and in GRB fireballs where a $\Gamma\approx 10^{2}-10^{3}$ 
is estimated.


\section{Numerical Method}

In these Monte Carlo codes we use both large angle and pitch angle scattering, assuming 
elastic scattering in the fluid frames and the important aim is to investigate the role 
of different scattering models in reference to spectral shape, acceleration time scales 
and angular distribution in different frames of reference with varying  shock obliquity 
at very high gamma plasma flows, ranging from $\Gamma=10-10^{3}$.
We consider oblique, sub-luminal and super-luminal 
shocks with $15^o\leq\psi\leq 75^o$ where $\psi$ is the angle between the shock normal 
and the magnetic field, seen in the shock frame.
The relativistic frames of references we use during the calculations
are, the local fluid frame (1,2), the normal shock frame (s) and
the de Hoffman-Teller frame (HT). In general, flow into and out of the shock discontinuity 
is not along the shock normal,  but a transformation is possible into the normal shock frame to render 
the flows along the normal (e.g. Begelman and Kirk, 1990) and for simplicity we assume such 
a transformation has already been made.\\  

\subsection{Sub-luminal case}
 
The Monte Carlo codes for the simulations in the sub-luminal case are similar to the 
treatment we followed in Meli and Quenby (2002a) for the parallel shock simulations, 
apart from the fact that more complicated relativistic formulas and different frames 
of references are used (which make the runs of the codes considerably more CPU-time consuming).
First, relativistic particles are injected upstream and are allowed to scatter in either 
fluid frame and across the shock discontinuity. We follow each particle trajectory across the shock 
according to the jump conditions and each particle leaves the system when it
'escapes'  far downstream at the spatial boundary or if it reaches a well defined maximum 
energy $E_{max}$, equal to $10^{10}E_{o}$. The spatial boundary is placed $\sim 100\lambda$
downstream in the large angle scatter case which proved an optimum number, 
but for small angle scattering, the possible dependence of the results on this parameter needed 
investigation (see the 'Results' section).
The compression ratio $r=\frac{V_1}{V_2}$ takes
the values of 4 and 3 for comparison with previous work with varying relativistic
Rankine-Hugoniot conditions and a similar version 
of Bednarz and Ostrowski (1998) splitting technique is used (see Meli and Quenby, 2002a)
so that when an energy is reached such that only a few accelerated particles remain, each particle 
is replaced by $N$ of statistical weight $1/N$ so as to keep roughly the same number of particles 
being followed. A guiding centre approximation is used, where the particle trajectory is followed in
two-dimensional, pitch angle and distance along {\bf B} space. The mean free path is calculated 
in the respective fluid frames by the formula: $\lambda=\lambda_{\circ} p_{1,2}$, 
assuming a momentum dependence to this mean free path for scattering along the field  and related to 
the spatial diffusion coefficient,
$K$, in the shock normal or $x$ direction by $K_{\|}=\lambda v/3$ and $K=K_{\|}cos^{2} \psi$.
Thus we specifically neglect cross-field diffusion, under conditions to be discussed in the next
paragraphs.\\ 
There are three conditions for the guiding centre approximation to be fulfilled 
relating to the ratio $K_{\perp}/K_{\|}$, the probability of hitting the shock at
the last scattering including the effect of field line wandering near the shock
and the across field jumps each scatter and we mention these, employing entirely
plasma frame variables. If 
$K_{\perp}/K_{\|}\leq0.1$, diffusion away from the shock is clearly predominantly 
along the mean field. Adapting quasi-linear theory for gyro-resonance scattering by a     
wave number, $k$, spectrum of fluctuations perpendicular to the mean field of power
$P(k)=Ak^{-n}$ we get (Jokipii, 1966),

\begin{equation}
K_{\|}=\frac{vB^{2}r_{g}^{2-n}}{\pi(2-n)(4-n)A}
\end{equation}

To match the usual assumption, $\lambda_{\|}\propto r_{g}$, we need $n$=1.
Then if $k_{\circ},k_{1}$ are the lower and upper limits to the wave number spectrum,
chosen to avoid divergence in fluctuation power,

\begin{equation}
\lambda_{\|}=\frac{r_{g}}{\pi}\frac{ln(k_{1}/k_{\circ})}{ \delta b^{2}}
\end{equation}

where $\delta b^{2}$ is the power in transverse fluctuations normalized to the mean field.
Perpendicular diffusion is likely to be predominantly due to field line wandering with 
(Jokipii and Parker, 1969), 

\begin{equation}
K_{\perp}=\frac{v}{4B^{2}}P_{\|}(k\rightarrow0)=
\frac{v}{4k_{\circ}}\frac{\delta b_{\|}^{2}}{ln(k_{1}/k_{\circ})}
\end{equation}

where $P_{\|}(k\rightarrow0)$ is the power in the parallel field at wave numbers near zero.
Even if the fluctuations are isotropic parallel and perpendicular to the field and
if $k_{\circ}^{-1}=L_{s}/2 \pi$ where $L_{s}$ is the scale size of the acceleration region,
with $\delta b\sim0.3$ and $k_{1}/k_{\circ}\sim10$, since at the highest reasonable $r_{g}$, 
$r_{g}\leq L_{s}/10$, we expect $K_{\perp}/K_{\|}\sim0.05$.  Hence at high $r_{g}$ the guiding centre
approximation for the mean field is reasonable. \\ 
At low $r_{g}$, perpendicular motion is entirely determined by the wandering of the field line bundle 
to which the particle is 'attached' and it is the local value of $\psi$, within a few $\lambda_{\|}$ 
of the shock, which determines the acceleration details.  
However, a second criterion that the guiding centre approximation remains reasonable is that 
$tan \psi \leq \lambda_{\|}/r_{g}$. This means physically that the gyrating particle does
not on average scatter far enough off from the original 'bundle' to
intersect the shock front, as it scatters at a distance of one parallel
mean free path from the front along the actual field line. The reason for the scattering, large angle 
or a reflection due to multiple small-angle scatterings, is not crucial. For the above parameters this 
criterion becomes $\psi\leq 84^{\circ}$. However, since the field itself 
'wanders' by an rms value $\sim 18^{\circ}$, the guiding centre approximation is probably limited to 
field-shock normal angles $\leq 66^{\circ}$, at all $r_{g}$.\\  
As explained by Baring et al. (1995), each scatter induces a jump across the field
which changes the distance to the shock by $\leq r_{g}$,  but this is unimportant in the numerical regime 
we consider where, $\lambda_{\|}\geq10r_{g}$.\\ 
The probability that a particle will move a distance $\Delta s$ along the field lines at pitch angle 
$\theta$ before it scatters, is given by the expression 
$Prob(\Delta s) \sim exp(-\Delta s/\lambda)$. The probability of finding 
a particle in a pitch angle interval $d\mu$ where $\mu=cos\theta$ is $\mu$,
and the time between collisions is $\upsilon |\mu|$ for particle velocity $\upsilon$ and $\mu$ 
chosen randomly between $-1$ and $+1$. Since cross field diffusion is neglected, 
$\Delta x=\Delta s cos \psi$.
To take into account increased downstream scattering, we take 
$\lambda_{\circ,1}=4\lambda_{\circ,2}$.\\  
In the shock rest frame, the flow velocity ($V_1$,$V_2$) for upstream and downstream respectively, is 
parallel to the shock normal and the magnetic fields ${\bf B}_1$ and  ${\bf B}_2$ are at an angle
$\psi_1$ and $\psi_2$ to the shock normal respectively. We adopt a geometry with $x$ in the flow 
direction, positive downstream, ${\bf B}_{1},{\bf B}_{2}$ in the $x-y$ plane and directed in the 
negative $x$ and $y$ directions and only ${\bf E}_{z}$ finite and in the positive $z$ direction.
A relativistic transformation is performed to the local fluid frame each time the
particle scatters across the shock.\\
Now, for the transformations between the shock frame and the de Hoffmann-Teller frame, we 
need to boost by a $V_{HT}$ speed along the shock frame where, $V_{HT}$  is equal to $V_1 tan\psi_1$.
The transformation to the de Hoffmann-Teller  frame though, is only possible if $V_{HT}$ is less or 
equal to the speed of light.  This means that $tan\psi_1\leq 1$ (sub-luminal case).
The particles ($\sim10^{5}$) of a weight  equal to 1.0, are injected far  upstream at a constant
energy of high gamma, which supposes that a pre-acceleration of the particles
has already taken  place, following for example the astrophysical scenario of
Gallant and Achterberg (1999) and as discussed in more detail in Meli and Quenby (2002a). 
The particles are left to move towards the shock, along the way they collide with the
scattering centers and consequently as they keep scattering between the upstream and 
downstream regions of the shock (its width is much smaller than the particle's gyroradius) 
they gain a considerable amount of energy (see figures of the next section).
For the case of highly relativistic flows, the definition of particle
pitch angle diffusion is that the scattering is limited within an upstream frame angle $\sim 0.1/\Gamma$ 
where, $\Gamma$ is the upstream gamma, measured in the shock frame (Gallant and Achterberg, 1999) and
investigated by Protheroe et al. (2002).
This condition is also included in the current Monte Carlo for appropriate comparisons. 
The energy of the particle is kept constant at the scattering centers and only the 
direction of the velocity vector is randomized. 
We put forward a condition for large angle scattering to occur
in Meli and Quenby (2002a), $r_{g,1}\leq \lambda_{1}/2 \Gamma_{1}^{2}$ 
where the mean free path and gyroradius within the scatter centre, 
$r_{g,1}$, are measured in the upstream frame and will also consider this case
as possible, but the reasoning,
both for the sub-luminal and the super-luminal situations is modified. 
However, we note Achterberg et al. (2001) exclude large angle scatterings in an extensive investigation 
of a variety of models.\\
Provided the field directions encountered are reasonably isotropic in the shock frame,
we are about to show $tan\psi_{1}=\Gamma_{1}^{-1}tan\psi_{s}\sim \Gamma_{1}^{-1}\sim \psi_{1}$
where '1' and 's' refer to the upstream and shock frames. Now, $V_{1}=c(1-0.5\Gamma_{1}^{-2})$
while, a particle at velocity $c$ in the upstream frame starting along the shock normal, needs 
a deflection of $\Gamma_{1}^{-1}$ radians to be reduced to $V_{1}$ resolved in the shock normal direction
and hence, be overtaken by the shock. This angle is the limit of gyration allowed in the upstream
field perpendicular to the shock normal before some large angle scatter must take place.
If $t_{u}$ is the time to the scatter and $\omega_{\perp}$ is the perpendicular gyrofrequency,
the angular deflection limit is $\omega_{\perp}t_{u}=\Gamma_{1}^{-1}$ where, $\omega_{\perp}=
u_{\perp}/r_{a,\perp,1}$ and $u_{\perp}=c\psi_{1}$. Then the allowed distance before large angle scatter
is $\lambda_{\|}=ct_{u}\leq r_{a,\perp,1}$. This means the Larmor 
radius of the high field 'bullet'
scatter centre is, $r_{g,1}\leq r_{a,1}/\Gamma_{1}^{2}$ using the previous condition, which requires 
field strength regimes as discussed in Meli and Quenby (2002a) for possible GRB situations.\\ 
Between the 'i th' and 'i+1 th' scattering, we check for a shock crossing up to down via ($c=1$ units)

\begin{equation}
\Delta x_{s}^{i+1}=\Gamma_{1}(\Delta x_{i}^{i}+V_{1}|\frac{\Delta x_{1}^{i} sec \psi_{1}}
{\upsilon_{1} cos\theta_{1}}|)
\end{equation}

\begin{equation}
\Delta t_{s}^{i+1}=\Gamma_{1}(|\frac{\Delta x_{1}^{i} sec\psi_{1}}{\upsilon_{1} cos \theta_{1}}|+V_{1}
\Delta x_{1})
\end{equation}

where $\psi_{1}$ is measured in the upstream frame. Defining ${\bf B}_{s}$ and ${\bf E}_{s}$ in the shock 
normal frame with  compression ratio $r$, we arrive at upstream frame quantities; 
$B_{1,x}=B_{s,x}$, $B_{1,y}=\Gamma_{1}(B_{s,y}+V_{1}E_{s,z})$, $E_{1,z}^{'}=0$ where 
$E_{s,z}=V_{1}B_{s} sin\psi_{1,s}$. Also $tan \psi_{1}=\Gamma_{1}^{-1} tan \psi_{s}$. 
Note the high relative velocity of the frames in the $x$ direction induces an electric field which via a back 
induction suppresses the $y$ component of B and swings the field  lines towards the $x$ axis
in the plasma frame. \\
Physically, for the occurrence of a shock crossing the conservation of the adiabatic 
invariant in the de Hoffman-Teller frame is used. However, to get into this frame we need all components 
of momentum and so need to assign a particle phase angle $\phi$ at random. Then, in $m=1$ units, 
the momentum components for particle mass $\gamma$ are,

\begin{equation} 
p_{1,x}=\gamma_{1}\upsilon_{1}cos\theta_{1}cos\psi_{1}-\gamma_{1}\upsilon_{1}sin\theta_{1}cos\phi_{1}sin\psi_{1}
\end{equation}

\begin{equation}
p_{1,y}=\gamma_{1}\upsilon_{1}cos\theta_{1}sin\psi_{1}+\gamma_{1}\upsilon_{1}sin\theta_{1}cos\phi_{1}cos\psi_{1}
\end{equation}

\begin{equation}
p_{1,z}=\gamma_{1}\upsilon_{1}sin\theta_{1}sin\phi_{1}
\end{equation}

A two stage transform is the initiated, first to the shock frame,

\begin{equation}
\gamma_{s}=\Gamma_{1}(\gamma_{1}+V_{1}p_{x,1})
\end{equation}

\begin{equation}
p_{s,x}=\Gamma_{1}(p_{x,1}+V_{1}\gamma_{1})
\end{equation}

\begin{equation}
p_{s,y}=p_{y,1}
\end{equation}

\begin{equation}
p_{s,z}=p_{z,1}
\end{equation}

A transformation is now made into the de Hoffmann-Teller frame with a boost along the negative $y$ axis with 
$V_{HT}=V_{1}tan\psi_{1}$. Then in this new frame, the $B$ components become 
$-B_{HT,x}=\Gamma_{HT}(-B_{s,x}-V_{HT}E_{s,z})$, $B_{HT,y}=B_{s,y}$ and hence 
$tan\psi_{HT,1}=\Gamma_{HT,1}tan\psi_{1}$. Here the field lines are swept towards the transformation 
velocity in the $y$ direction under frame transform. Energy and the $y$ component of momentum are then 
transformed into the de Hoffmann-Teller frames according to, $\gamma_{HT}=\Gamma(\gamma_{s}+V_{HT}p_{s,y})$ 
and $p_{HT,y}=\Gamma_{HT}(p_{s,y}+V_{HT}\gamma_{s})$  allowing the particle pitch angle to be obtained 
from $cos\theta_{HT}={\bf p}\cdot{\bf B}/pB$.  Then the particle that crosses the shock from upstream, is 
transmitted only if its pitch angle is less than the critical pitch angle:
 
\begin{equation}
\theta_{c}=arcsin\sqrt{\frac{{\bf B}_{HT,1}}{{\bf B}_{HT,2}}}
\end{equation}

From the conservation of the first adiabatic invariant we can find the new 
pitch angle in the downstream frame and similar transformations allow the particle scattering to 
be followed in this frame.

\subsection{Super-luminal case}

For the super-luminal case a variation of the computational model is applied, compared
to the one depicted above. For this investigation we employ once again a Monte 
Carlo scheme and consider the motion of a particle of momentum $p$ in the magnetic
field {\bf B}. As we have mentioned, in the super-luminal situation it is not possible 
to transform to a single frame where {\bf E}=0 (de Hoffmann-Teller frame). Hence, for our 
investigation the most convenient frames of reference to use are the fluid frames 
(where still the electric field is zero) and the shock frame which we will use only 
as a frame to check whether upstream or downstream conditions apply. 
We inject $\sim 8\times 10^{5}$ particles in order to keep reasonable statistics
throughout the simulations. We consider large angle scattering, which is 
calculated in the respective fluid rest frames, based on the possible situation previously discussed.
Also Bednarz and Ostrowski (1996) find that only in the large angle scattering case does
a power law develop and thus this is the most favourable case to further investigate. It will at least
yield a limit to the acceleration a super-luminal shock may provide.   
Initially we inject the particles $100\lambda$ from the shock and we follow  their guiding 
center in the upstream frame as in the sub-luminal case
until  after the appropriate transformation to the shock frame the particle
reaches the shock  at $x_{sh}=0$. At this juncture, there is no easy approximation to 
determine the probability of shock crossing or reflection. We change the model to following a 
helical trajectory, in the fluid frames upstream ('1') or downstream ('2') respectively, 
where the velocity coordinates of the particle are calculated in a three dimensional space as follows:
 
\begin{equation}
\upsilon_{x_1}=\upsilon_1 cos\theta_1 cos\psi_1-\upsilon_1 sin\theta_1 cos\phi_1 sin\psi_1,
\end{equation}
 
\begin{equation}
\upsilon_{y_1}=\upsilon_1 cos\theta_1 sin\psi_1+\upsilon_1 sin\theta_1 cos\phi_1 cos\psi_1
\end{equation}
 
and,
 
\begin{equation}
\upsilon_{z_1}=-\upsilon_1sin\theta_1 sin\phi_1
\end{equation}
 
where $\theta_1$ is the pitch angle and $\psi_1$ is the angle between the magnetic field
and the shock normal.
 
We follow the trajectory in time, using $\phi_1=\phi_{\circ}+\omega t$, where
$\phi_1\in(0,2\pi)$ and $t$ is the time from detecting shock presence at $x_{sh}$,
$y_{sh}$, $z_{sh}$ by using,
 
\begin{equation}
dx=x_{sh}+\upsilon_{x_1} \delta t
\end{equation}
 
\begin{equation}
dy=y_{sh}+\upsilon_{y_1} \delta t
\end{equation}
 
and,
 
\begin{equation}
dz=z_{sh}+\upsilon_{z_1} \delta t
\end{equation}

assuming that $\delta t=r_{g}/Hc$, where $r_{g}$ is the Larmor radius,  $H \ge 100$.
The particle's gyrofrequency $\omega$ is given by the relation:
 
\begin{equation}
\omega_1=\frac {e |{\bf B}_1|}{\gamma_1}
\end{equation}
 
${\bf B}_1$ is the magnetic field, $\gamma_1$ is the particle's
gamma and $e$ is its charge in gaussian units. For a matter of convenience
though, the last relation is  transformed in units of $c/sec$.
All the above relations apply to the downstream case by only changing respectively
the signs from 1 to 2, and all the calculations are performed in the upstream or
downstream frames. Because of the peculiar properties of the helix we need to establish 
where a particle -in the upstream frame- of a particular $\theta,\phi$ \textit{first} 
encounters the shock. To establish when this happens, we choose to go back a whole period,
 
\begin{equation}
T_1=\frac{2\pi}{\omega_1}
\end{equation}
 
by reversing signs of the helix velocity coordinates and keep checking throughout the
simulation to see if the particle's trajectory encounters the shock again. The starting 
point for transforming to the downstream frame is the \textit{furthest} upstream shock crossing.
After making the suitable relativistic transformations to the downstream fluid frame by
calculating the ($x_2,y_2,z_2,t_2$) and ($\upsilon_{x_2},\upsilon_{y_2},\upsilon_{z_2}$)
coordinates, the momentum  $P_2$ and the gamma $\gamma_2$ of the particle, we follow the
trajectory of the particle for a whole downstream period,
 
\begin{equation}
T_2=\frac{2\pi}{\omega_2}
\end{equation}
  
checking to see whether the particle meets the shock again, by transforming to the
shock frame. If the particle meets the shock then the suitable transformations to 
the upstream frame are made again and we follow the particle's trajectory as described above.
If the particle never meets the shock, than its guiding center is followed, the same 
way as mentioned earlier for the upstream side after the injection and it is 
left to leave the system if it reaches a well defined $E_{max}$ momentum boundary or,
a spatial boundary of $100\lambda$.


\section{Results}

In this section we present results of the simulations described above.
We show similar figures regarding the acceleration time decrease, 
mean energy gain, spectral shapes and angular distributions for both sub- and super-luminal situations.
 First, the code runs have been tested in the non-relativistic and 
mild relativistic regime using  $V_{1}\sim 0.1c-0.6c$ and are found to be in accordance 
with non-relativistic theory and mild relativistic theoretical predictions 
(e.g. Kirk and Schneider, 1987a,b). In the highly relativistic regime for the sub-luminal case, 
the Lorentz factors of the flow used are equal to 10, 50, 200, 500 and 990. 
It is presumed that the injected particles are already relativistic with gammas just higher than 
the flow gammas. The compression ratio, $r$, has values of 3 and 4.
The value of $r$=3 is expected to mimic the physical condition of an ultra relativistic
gas. The value $r$=4 corresponds to the extreme  hydrodynamic limit, where the magnetic 
field pressure is unimportant, but the shock speed in the upstream frame tends to $c$. However we will see 
from the figures that the value 3 or 4  does not alter the results (possibly
due to the mildly-relativistic nature of the downstream plasma).

\begin{figure}[h!]
\begin{center}
\epsfig{figure=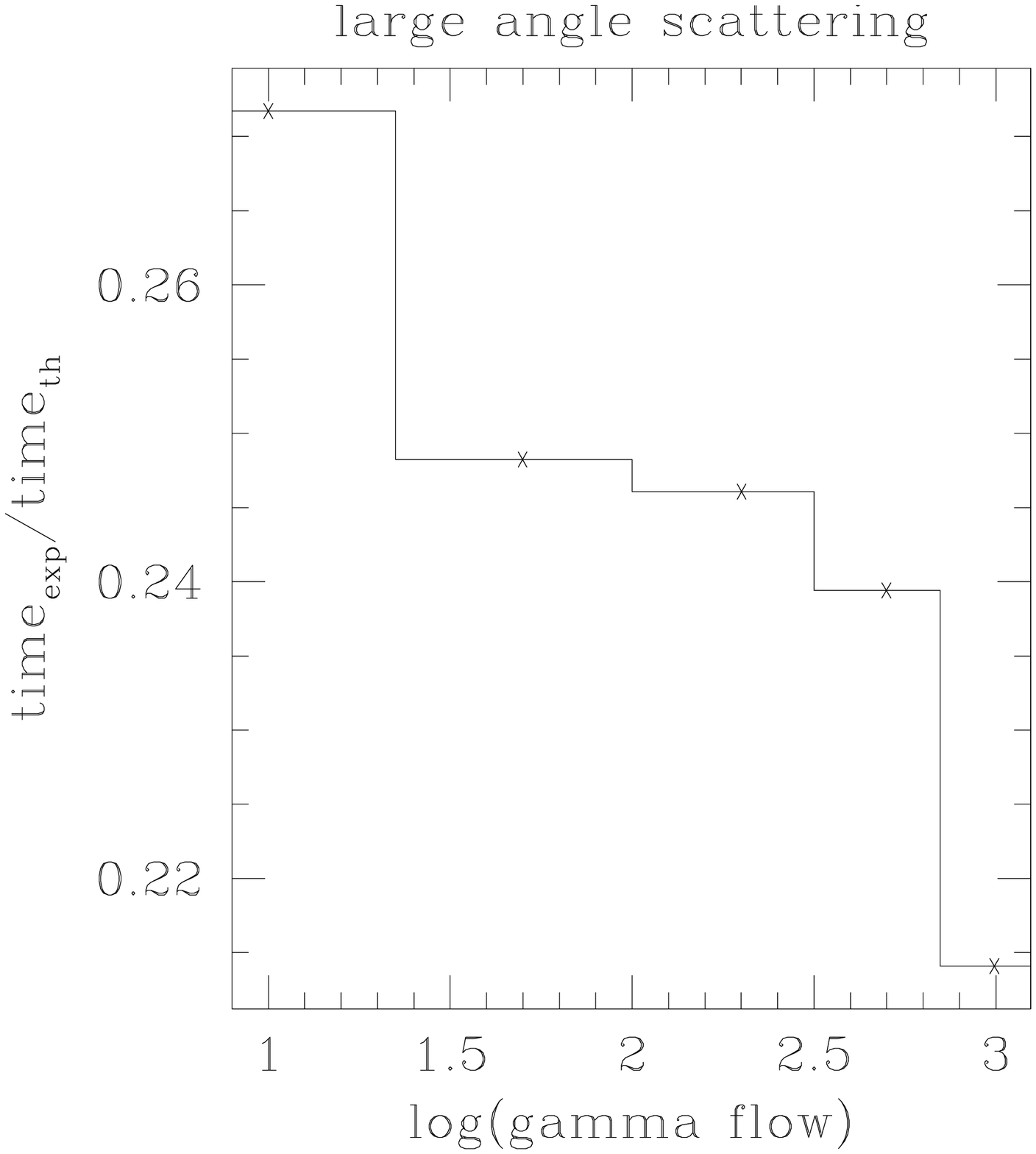, width=5.0cm}
\epsfig{figure=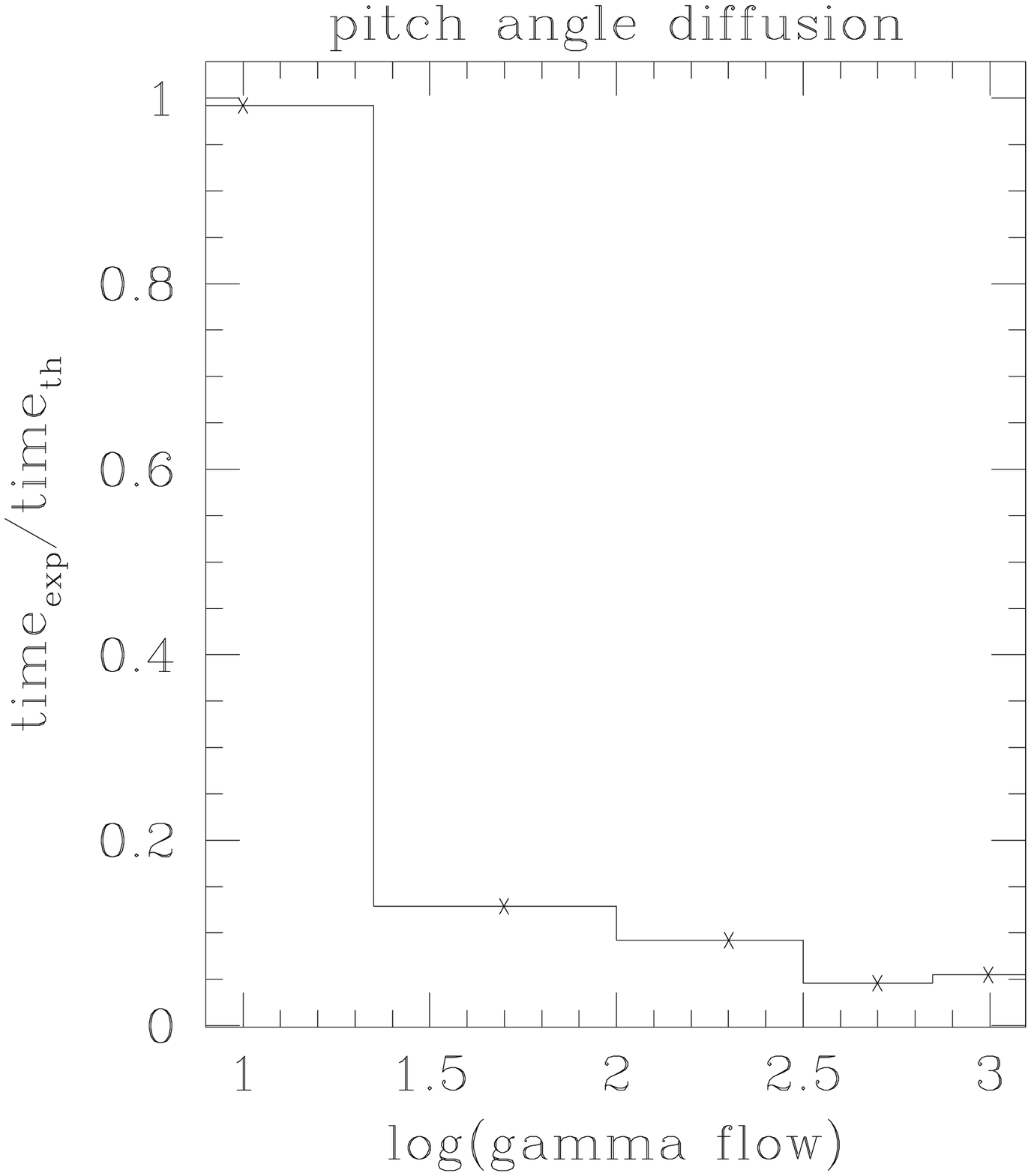, width=5.0cm}
\end{center}
\end{figure}

\begin{figure}[h!]
\begin{center}
\epsfig{figure=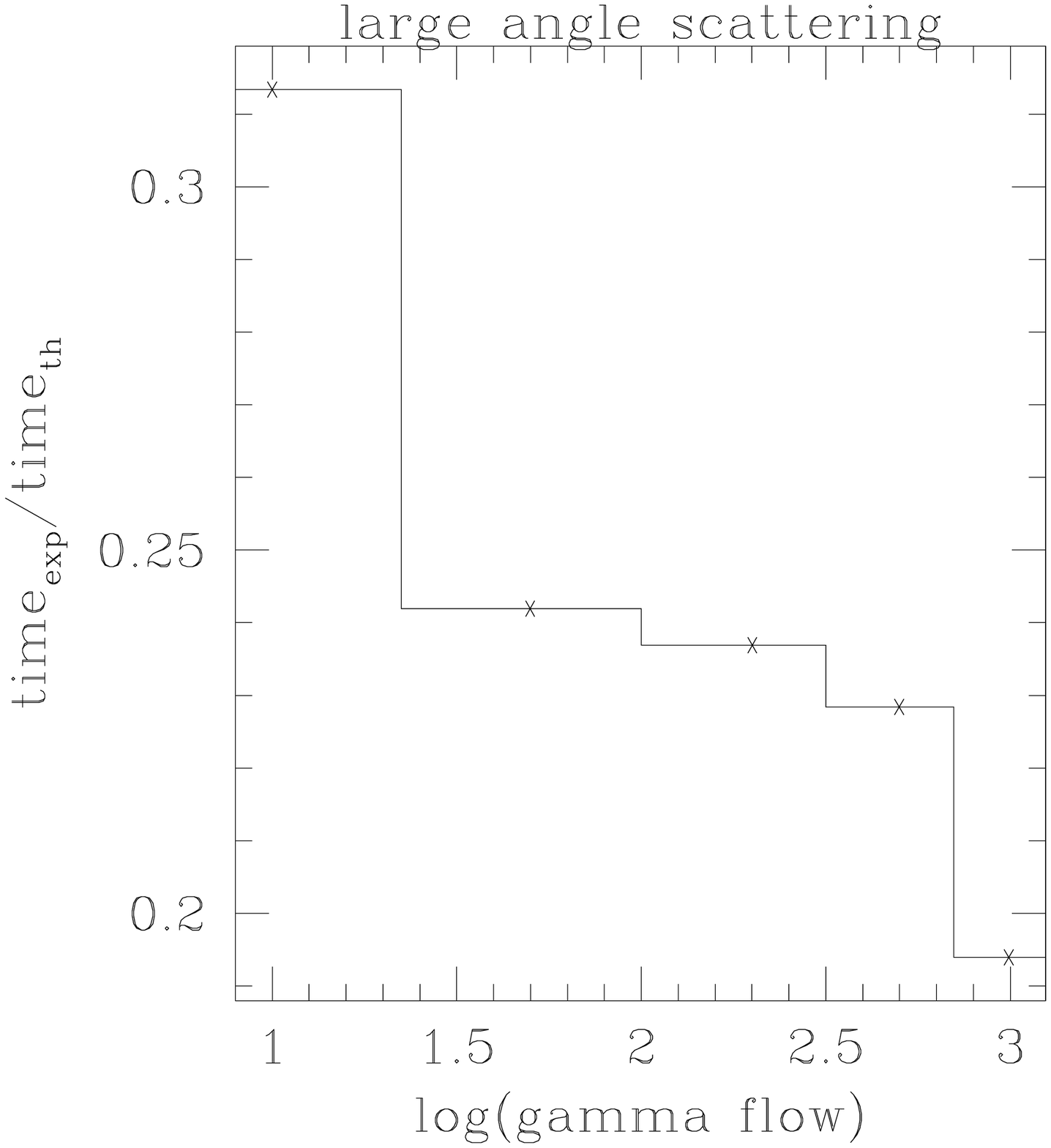, width=5.0cm}
\epsfig{figure=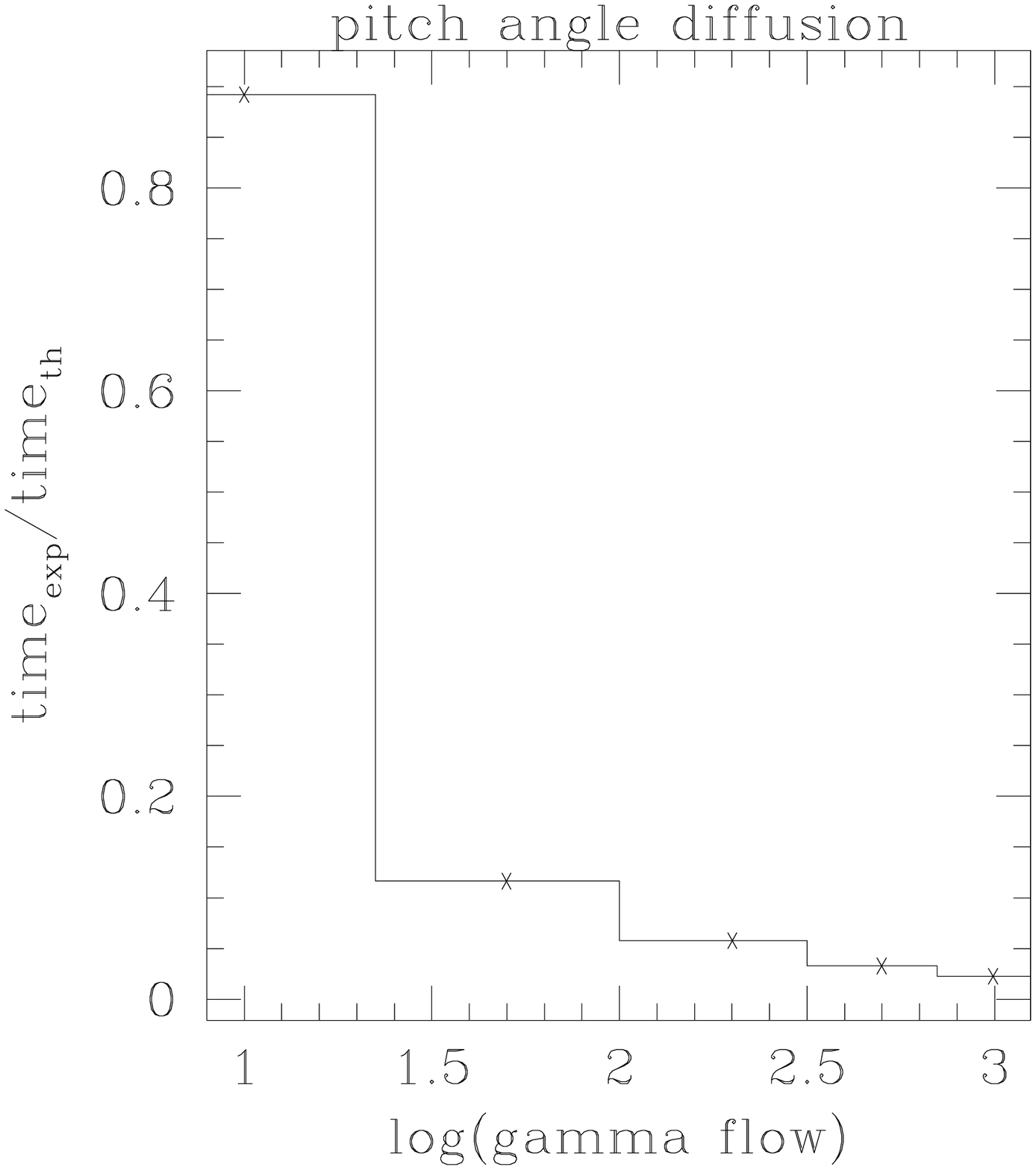, width=5.0cm}
\caption{The ratio of the computational time to the theoretical
acceleration time constant versus the logarithm of the upstream gamma flow.
Measurements are made at the  de Hoffmann-Teller frame. The top plots show the acceleration 
time decrease in the case of the large angle scattering (left), pitch angle diffusion (right), 
for $\psi_1=15^{\circ}$, while the bottom ones are for $\psi_1=35^{\circ}$.}
\end{center}
\end{figure}

\begin{figure}[t!]
\begin{center}
\epsfig{figure=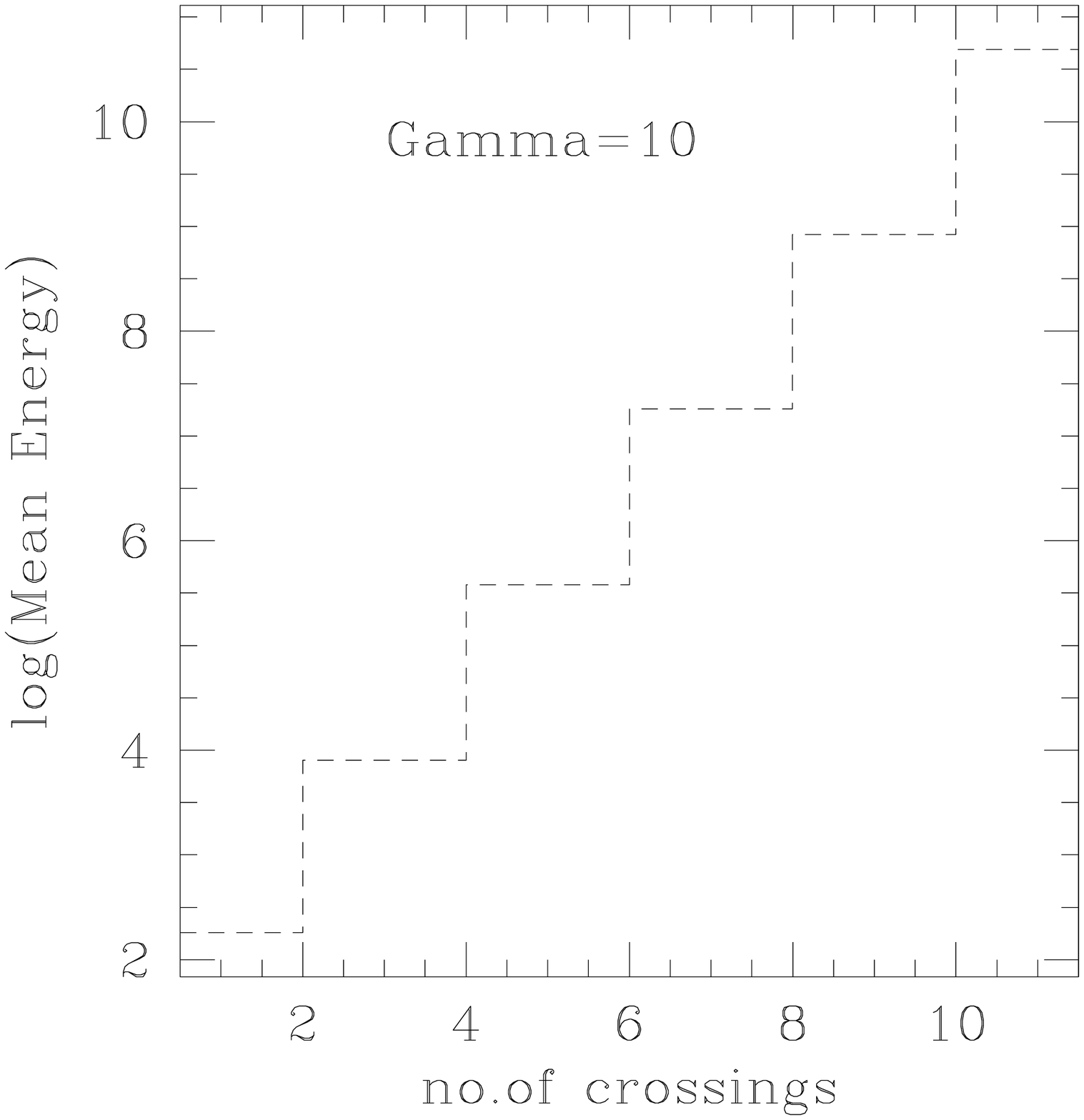, width=5.0cm}
\epsfig{figure=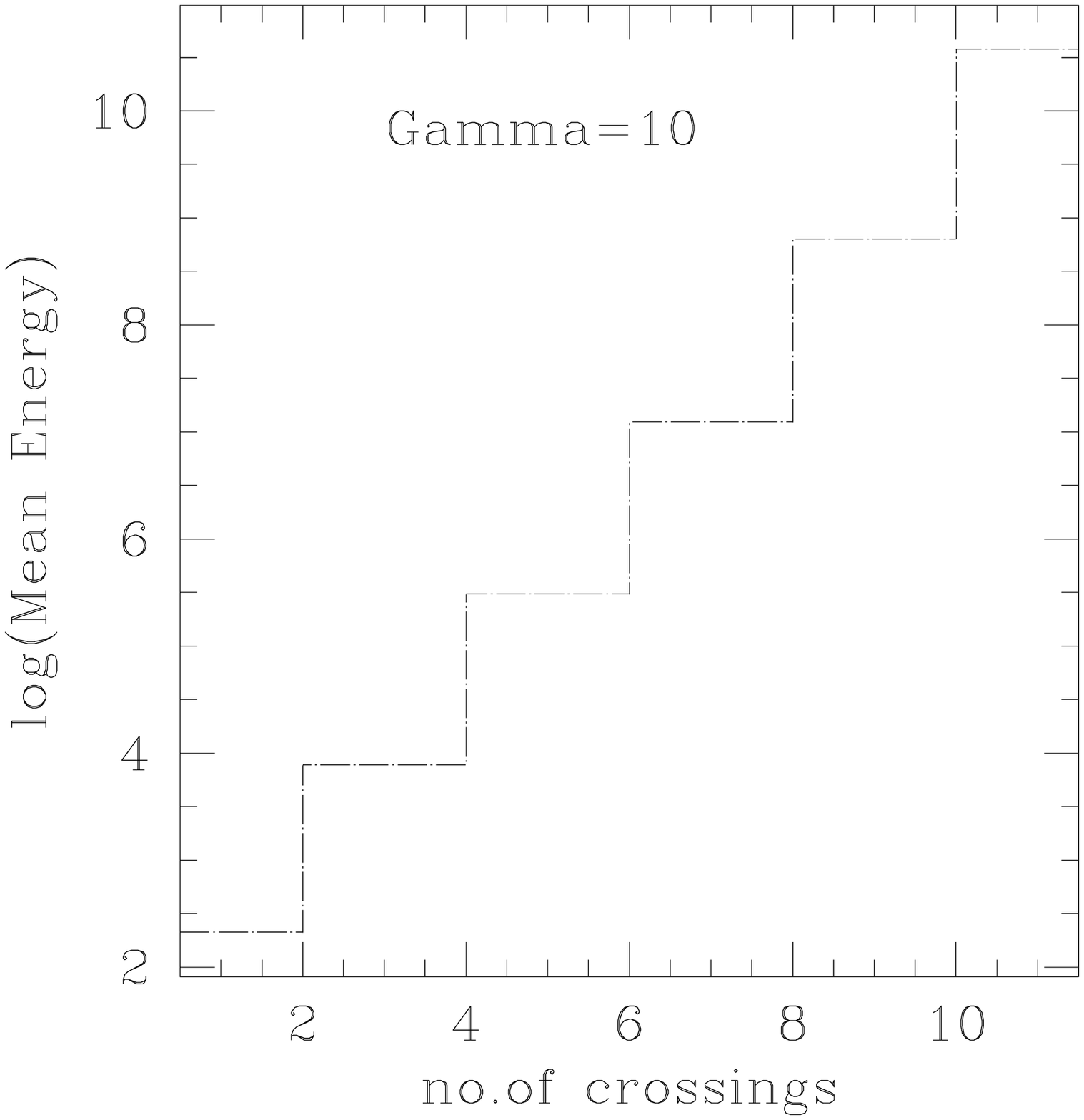, width=5.0cm}
\end{center}
\end{figure}

\begin{figure}[h!]
\begin{center}
\epsfig{figure=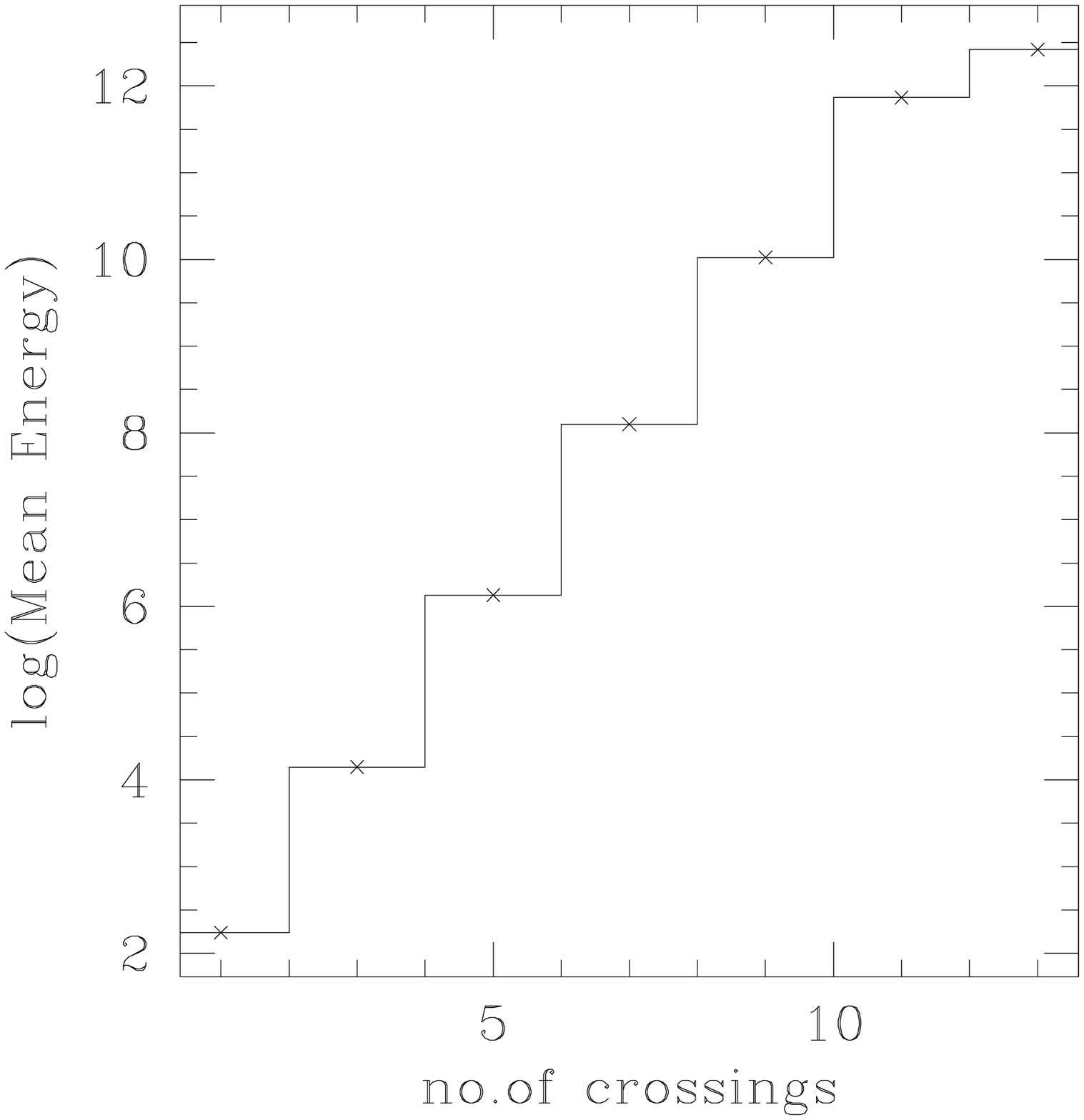, width=5.0cm} 
\epsfig{figure=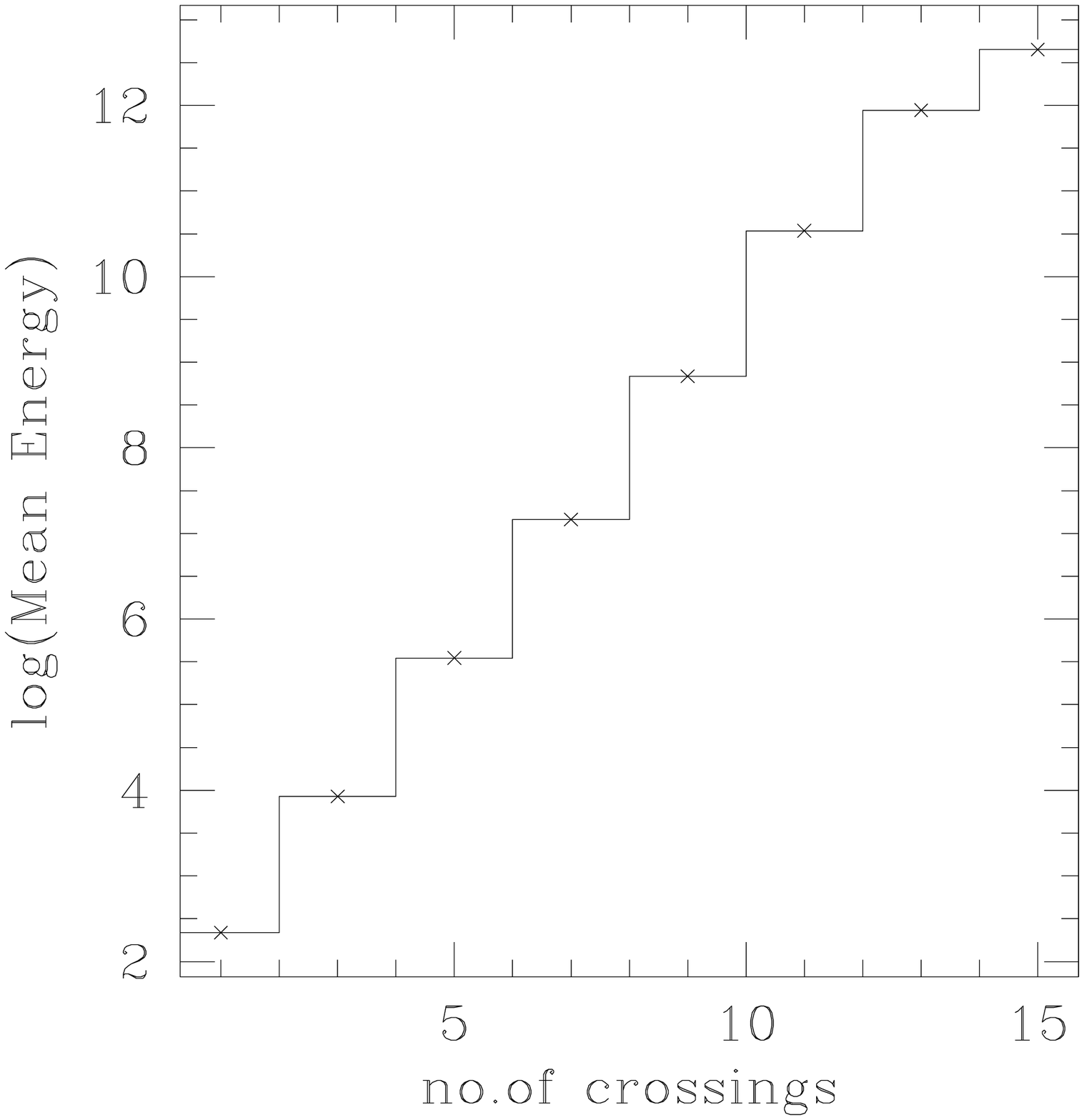, width=5.0cm}
\caption{The logarithm of the mean-energy gain of particles versus the number of  shock crossings 
(1-3-5-7-9-11) for $\Gamma$=10 and $r$=4.
Top plots: Large angle scattering,  $\psi$=15$^{\circ}$ (left),
$\psi$=35$^{\circ}$ (right).
Bottom plots: Pitch angle diffusion, $\psi$=15$^{\circ}$ (left),
$\psi$=35$^{\circ}$ (right).  We observe the $\Gamma^2$ gain factor between the 1$^{st}$ and 3$^{rd}$ 
shock crossing (one cycle) and the efficient energy multiplication in subsequent crossings.}
\end{center}
\end{figure}

Two sub-luminal inclination cases are studied, equal to $15^{\circ}$ and $35^{\circ}$. Particles 
are injected isotropically at 100$\lambda$ upstream in the shock frame and when they reach the 
spatial boundary placed at 100$\lambda_{\|}$ downstream they exit the simulation for the large angle case.
We apply large angle and pitch angle scatters. For pitch angle scatter, the stability of the results against 
boundary position enabled a suitable downstream exit to be chosen (1000$\lambda$) where here, $\lambda$ is the
distance between scatterings. The mean free path ($\lambda$), 
is chosen to have the value of $\lambda\sim 20r_{g}$, close to the value that 
Quenby and Lieu (1989) used, based on a summary of interplanetary transport simulations referring to 
the only astrophysical plasma where we have any detailed knowledge of the magnetic field properties
and performed by Moussas et al., 1992 (and references therein), but in most investigations it is
only the ratio of up/down stream mean free paths and their ratio to the spatial boundary
which matters.
The momentum boundary is equal to $E_{o}10^{14}$, where $E_{o}$ is equal to the 
initial energy of the particle at the injection.
\begin{figure}[h!]
\begin{center}
\epsfig{figure=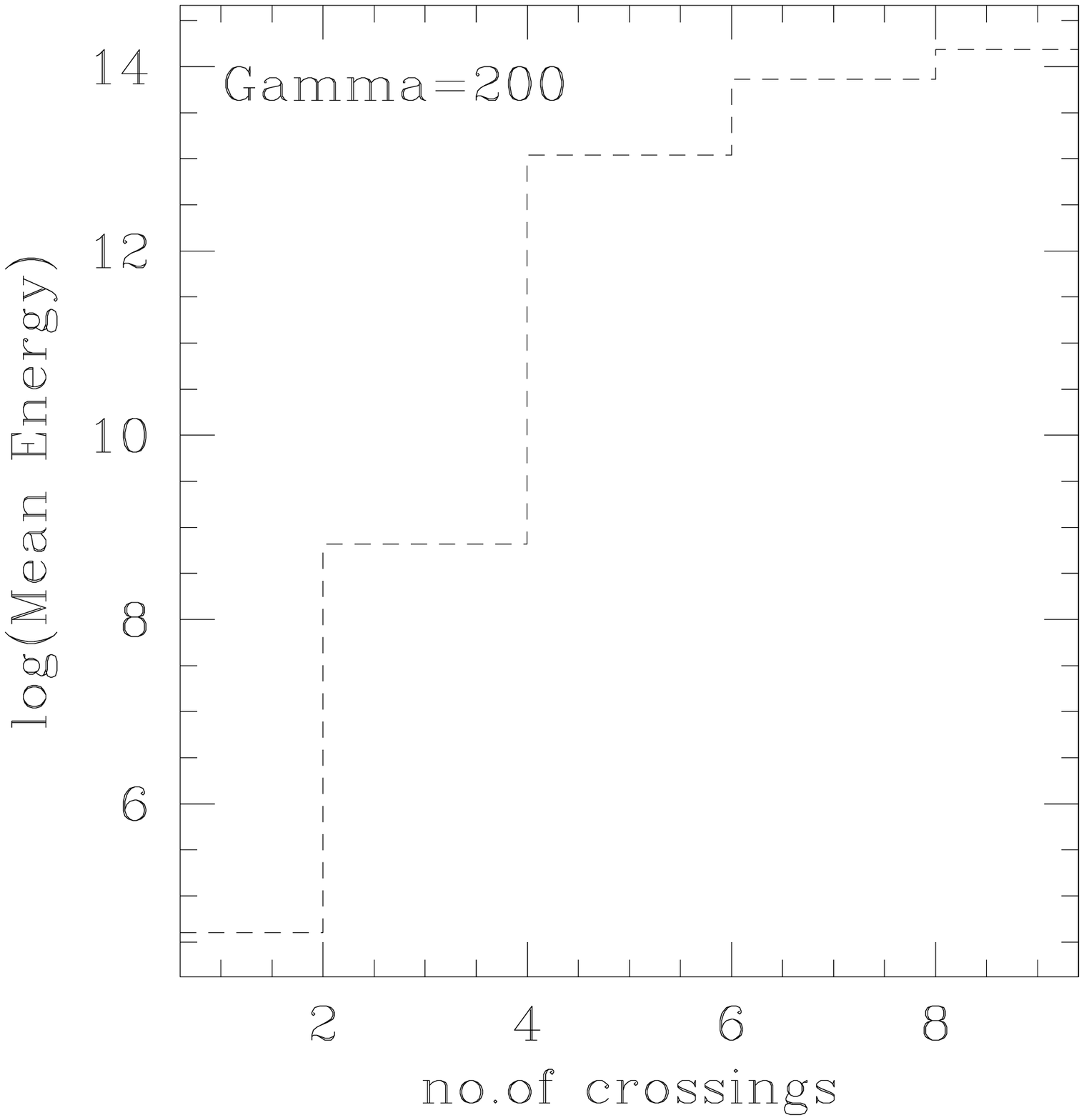,width=5.0cm}
\epsfig{figure=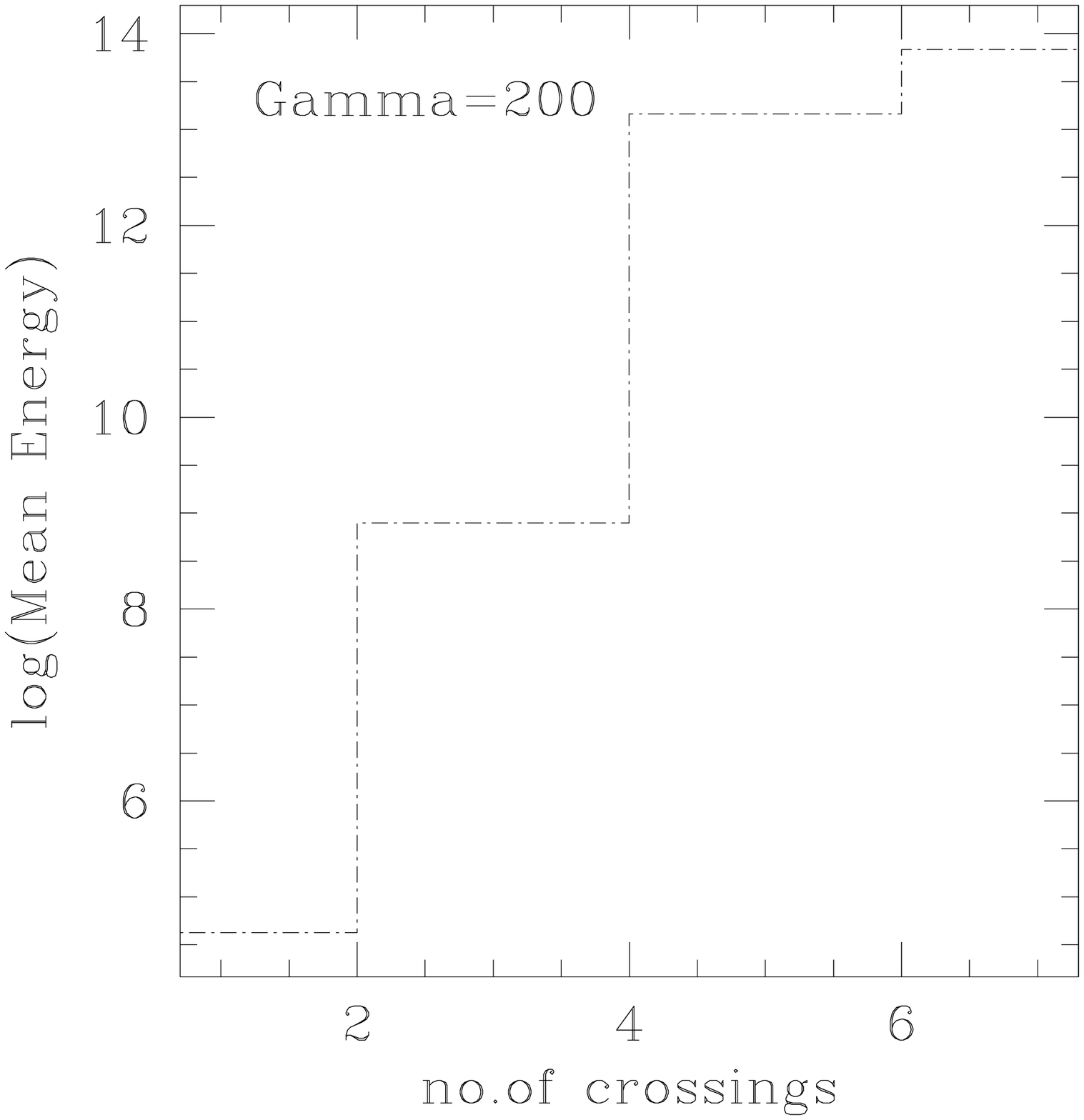,width=5.0cm}
\end{center}
\end{figure}
\begin{figure}[h!]
\begin{center}
\epsfig{figure=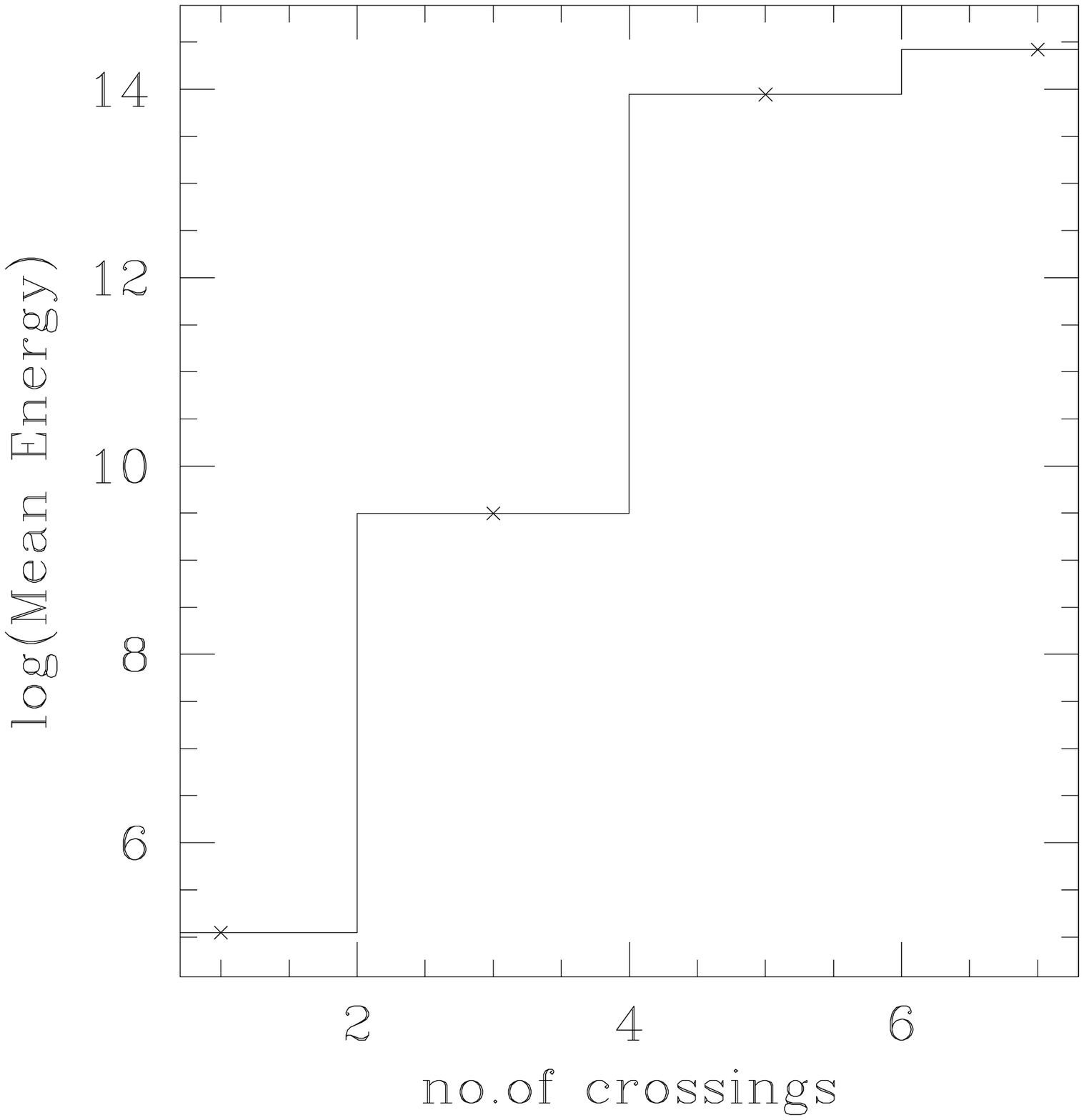,width=5.0cm}
\epsfig{figure=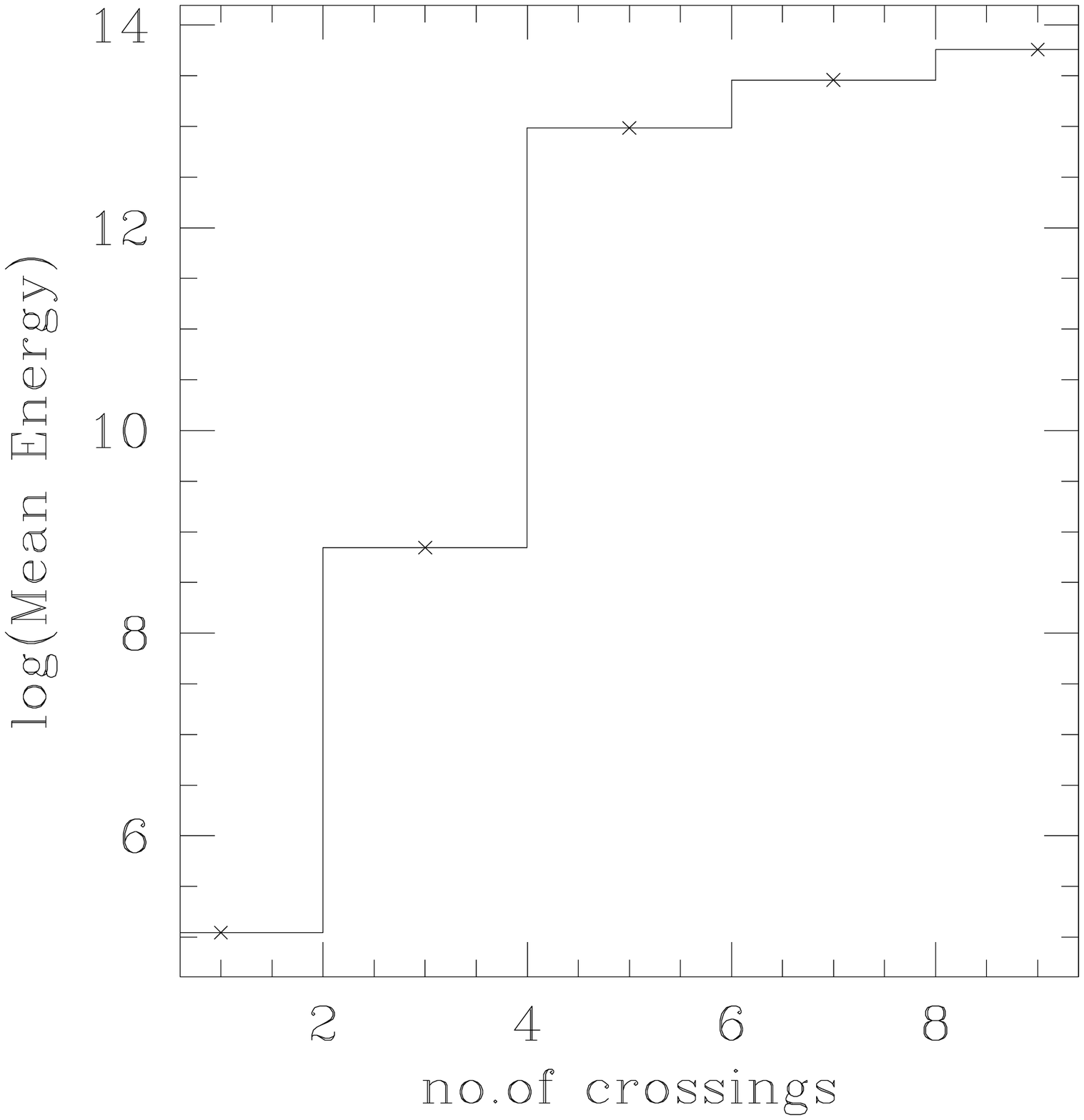,width=5.0cm}
\caption{The logarithm of the mean-energy gain of the particles
versus the number of shock crossing (1-3-5-7) for $\Gamma$=200.
Top  plots: Large angle scattering, $\psi$=15$^{\circ}$, $r$=4 (left),
$r$=3 (right). Bottom plots: Pitch angle diffusion, $\psi$=35$^{\circ}$, $r$=4 (left),
$r$=3 (right). We observe the $\Gamma^2$ gain factor between the 1$^{st}$ and 3$^{rd}$ shock crossing 
and the efficient energy multiplication in subsequent crossings.}
\end{center}
\end{figure}
For oblique sub-luminal shocks, we show in figure 1 the ratio of
computational acceleration time to the analytic non-relativistic time constant,
versus the logarithm of the upstream flow gammas used. The measurements are all made in the de Hoffmann-Teller
frame as in Lieu et al. (1994) where, it was found that this was a relevant frame to observe 
acceleration 'speed-up'.
Then the non-relativistic acceleration time is,

\begin{equation}
\tau=\frac{3}{V_{1}cos\psi_{1}-V_{2}cos \psi_{2}}\left(\frac{K_{n,1}}{V_{1}cos\psi_{1}}-\frac{K_{n,2}}{V_{2}
cos\psi_{2}}\right)
\end{equation}  

When considering pitch or small angle scattering, the value of $\lambda_{||}$ employed is that for 
$\sim\pi/2$ scatter after random pitch change so that, $\lambda_{\|}=(\pi^{2}/4)(\lambda/\delta \theta^{2})$
where $\lambda$ is the distance between $\delta \theta$ collisions. We observe a significant acceleration
time decrease of a factor $\sim$ 5-10, which is comparable to that of the parallel 
shock 'speed up' found in Ellison et al. (1990) and Meli and Quenby (2002a).
Basically, this 'speed up' effect observed is due to the $\Gamma_{1}$ energy increase each shock crossing, 
$\Gamma_{1}^{2}$ per cycle although it can be reduced by the very small pitch angles the particle 
may have in the small angle case as it 
crosses the shock upstream to downstream. 
In figures 2-4  the mean energy gain per crossing is presented for large angle
scattering and pitch angle diffusion, with $\psi_{1}$ equal to $15^{\circ}$, $35^{\circ}$, 
$r$=4 and $r$=3  and $\Gamma=$10, 200 and 990. Particle energy is measured in $\gamma$ units 
and so the plots are independent of mass and sign of charge. The relative independence on $r$ but 
the similarity in the trend from $\Gamma^{2}$ to $\Gamma$ in energy dependence and finally 'saturation' 
found in the parallel case (Meli and Quenby, 2002a) are seen. The gain per crossing falls off 
noticeably faster as $\Gamma_{1}$ increases.
\begin{figure}[h!]
\begin{center}
\epsfig{figure=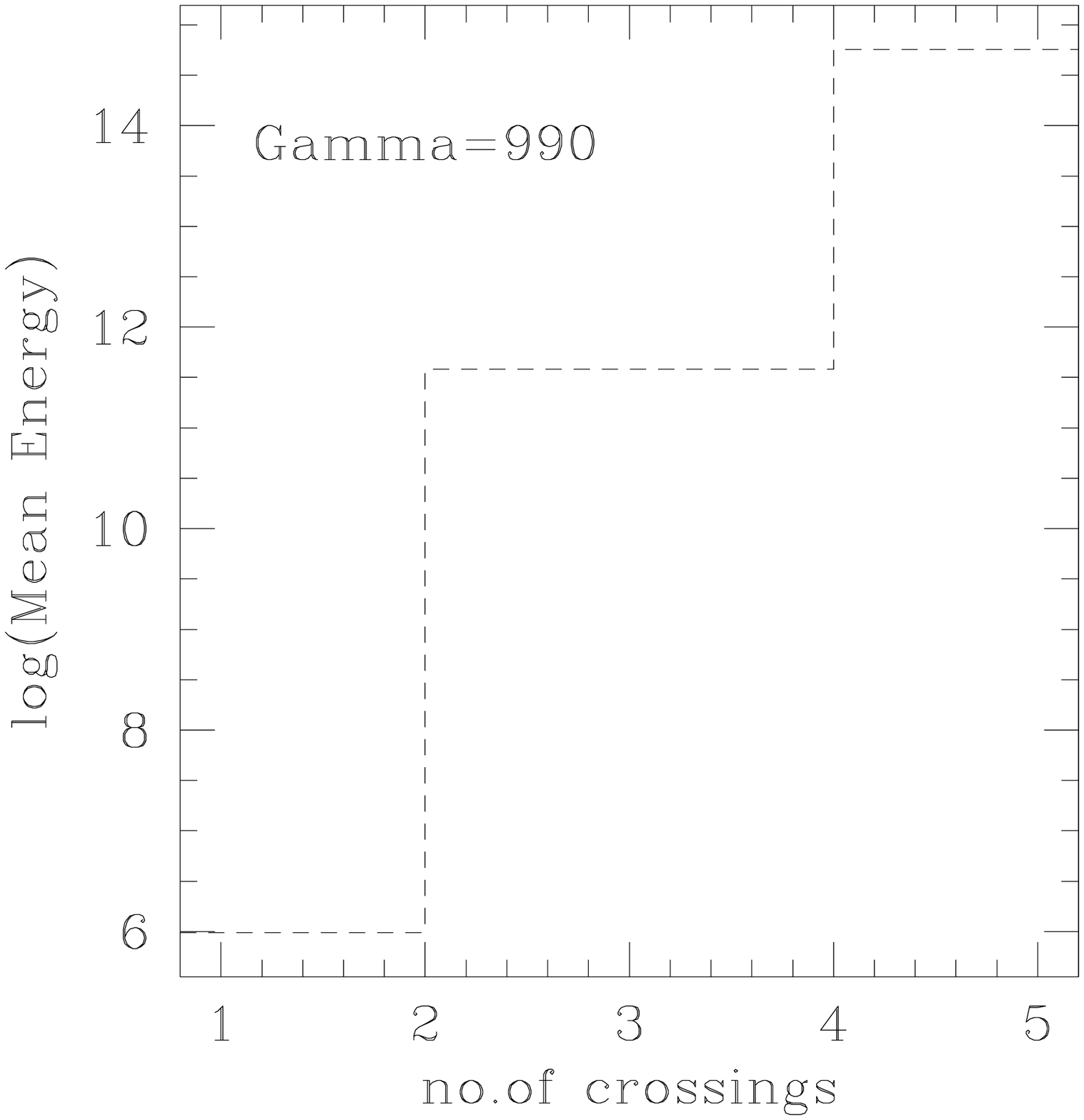,width=5.0cm}
\epsfig{figure=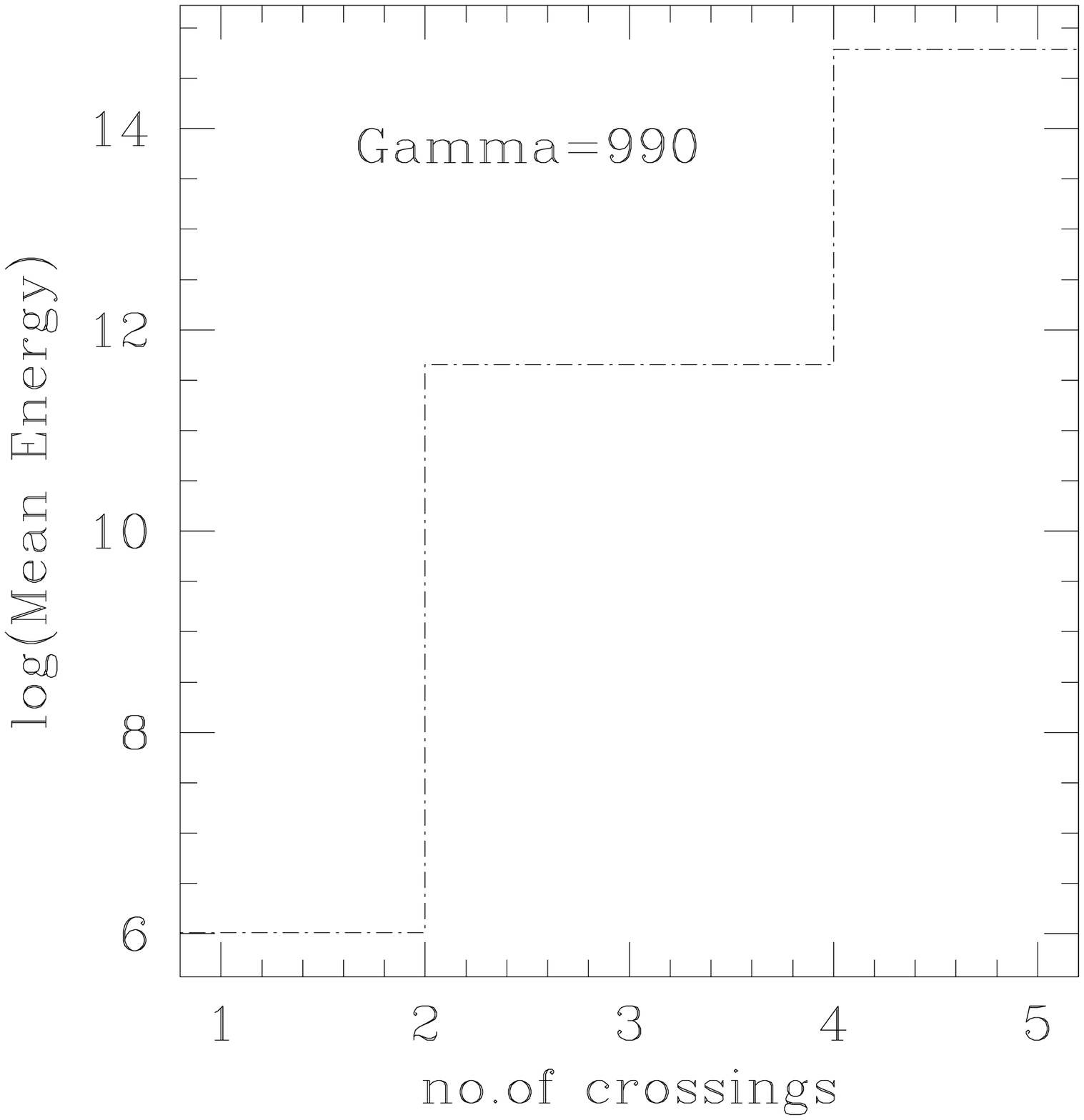,width=5.0cm}
\end{center}
\end{figure}

\begin{figure}[h!]
\begin{center}
\epsfig{figure=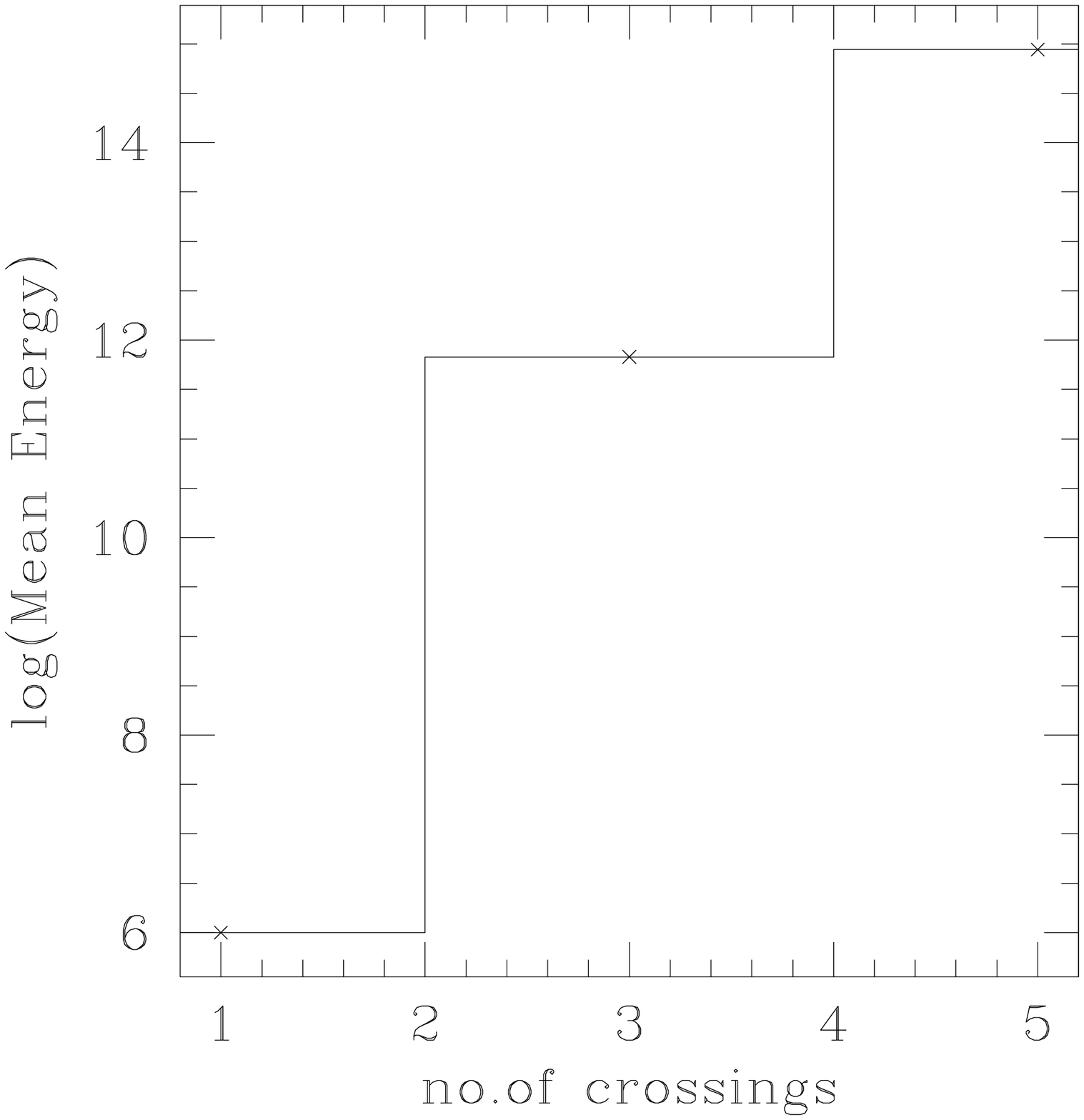,width=5.0cm}
\epsfig{figure=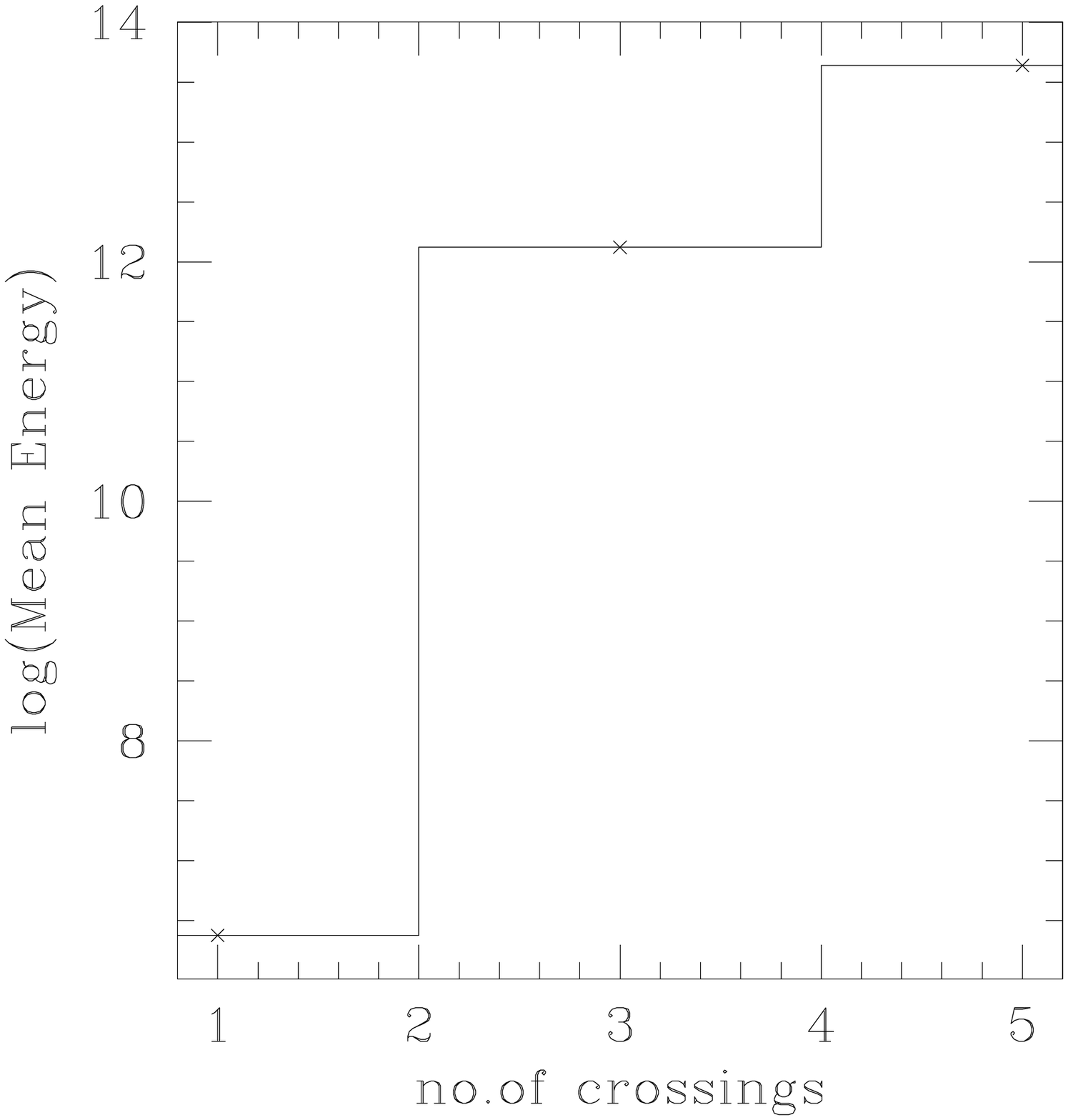,width=5.0cm}
\caption{The logarithm of the mean-energy gain of the particles
versus the shock crossing (1-3-5) for $\Gamma$=990. Top plots: Large angle scattering, $\psi$=15$^{\circ}$ 
and $r$=4 (left), $r$=3 (right). Bottom plots: Pitch angle diffusion,  $r$=4, $\psi$=15$^{\circ}$ (left), 
$\psi$=35$^{\circ}$ (right).}
\end{center}
\end{figure}
It must be emphasized that a full model would include loss processes which, would limit the maximum 
$\Gamma$ obtainable. Note that proton energies $\sim 10^{20}$ eV can be reached in all 
models computed before the saturation sets in, provided faster loss mechanisms are not present. 
This is despite the relatively small amount of upstream deflection allowed before particles are 
swept downstream on all cycles except the first which, limits the first order 'Fermi' gain available 
(see Gallant and Achterberg, 1999, where the full $\Gamma_{1}^{2}$ factor requires isotropization in each frame). 
Rather different spectra shapes are seen in figures 5-8 for $\Gamma$=10, 50, 500 and 990, depending 
on whether it is pitch angle or large angle scattering, at
$\psi_{1}=15^{\circ}$ and $\psi_{1}=35^{\circ}$. The sensitivity to downstream boundary position
has been checked up to a distance $r_{b}=2 \cdot 10^{5}\lambda$ where, $\lambda$ is distance between 
small angle scattering for this model.
\begin{figure}[t!]
\begin{center}
\epsfig{figure=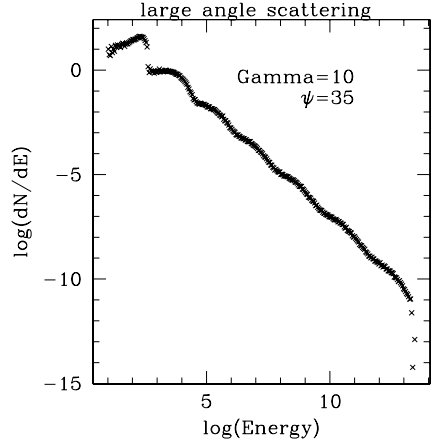,width=6.0cm}
\caption{Spectral shape for upstream gamma flow equal to 10, large angle scattering, 
$r$=4 and $\psi$=35$^{\circ}$. We observe the smoothness of the spectral shape compared to 
larger upstream gammas.}
\end{center}
\end{figure}

\begin{figure}[h!]
\begin{center}
\epsfig{figure=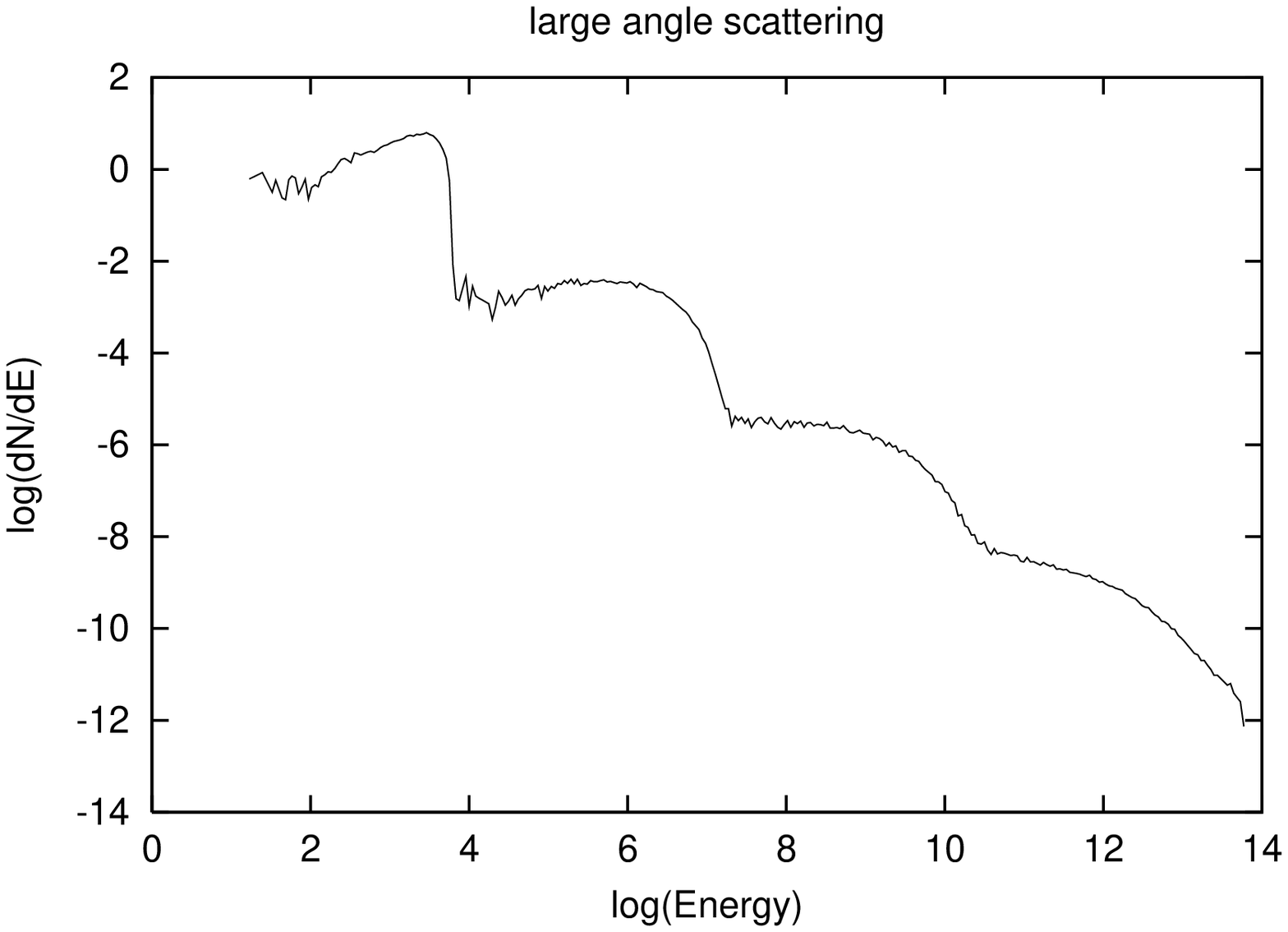,width=6.8cm}
\epsfig{figure=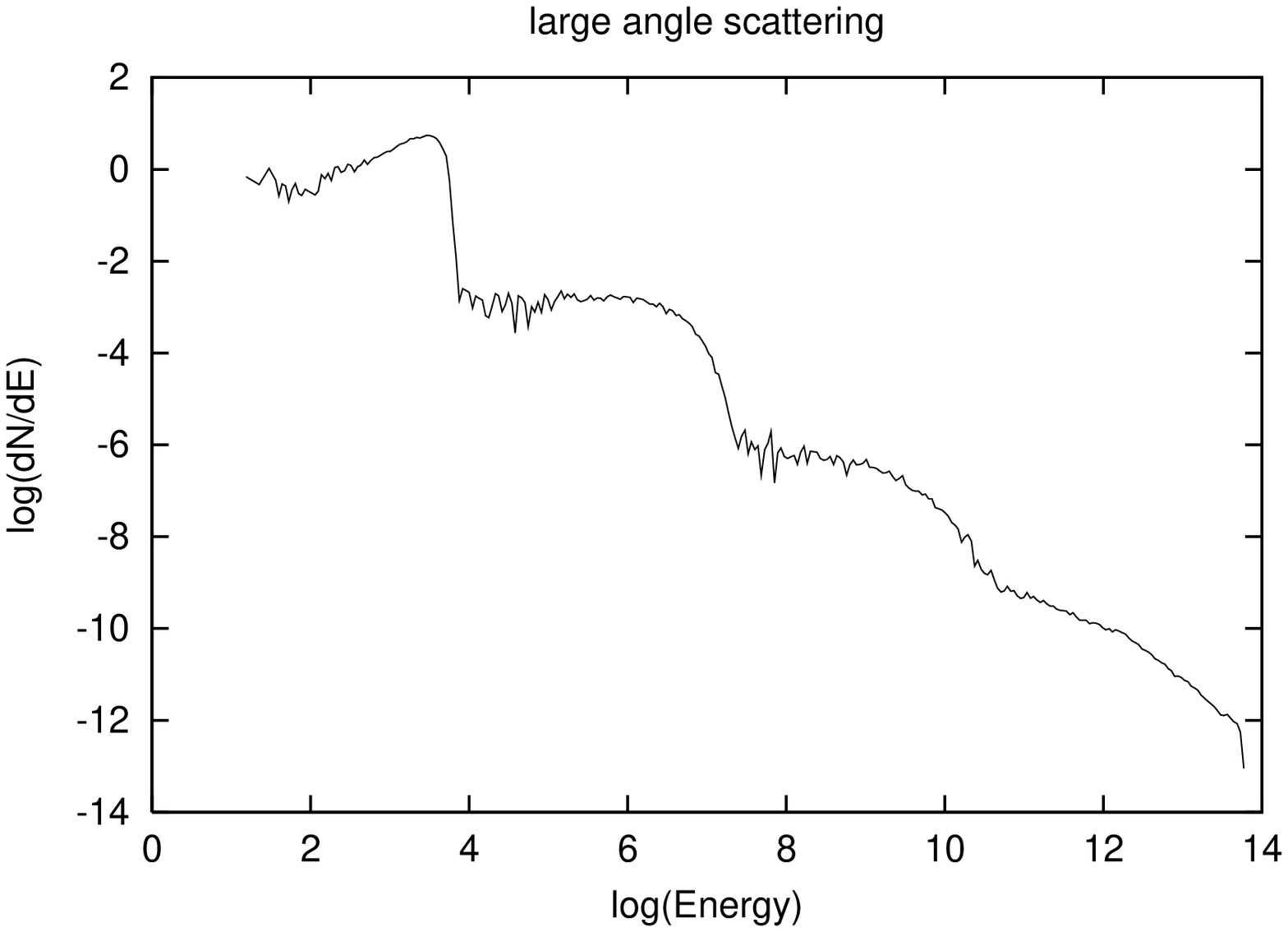,width=6.8cm}
\end{center}
\end{figure}

\begin{figure}[h!]
\begin{center}
\epsfig{figure=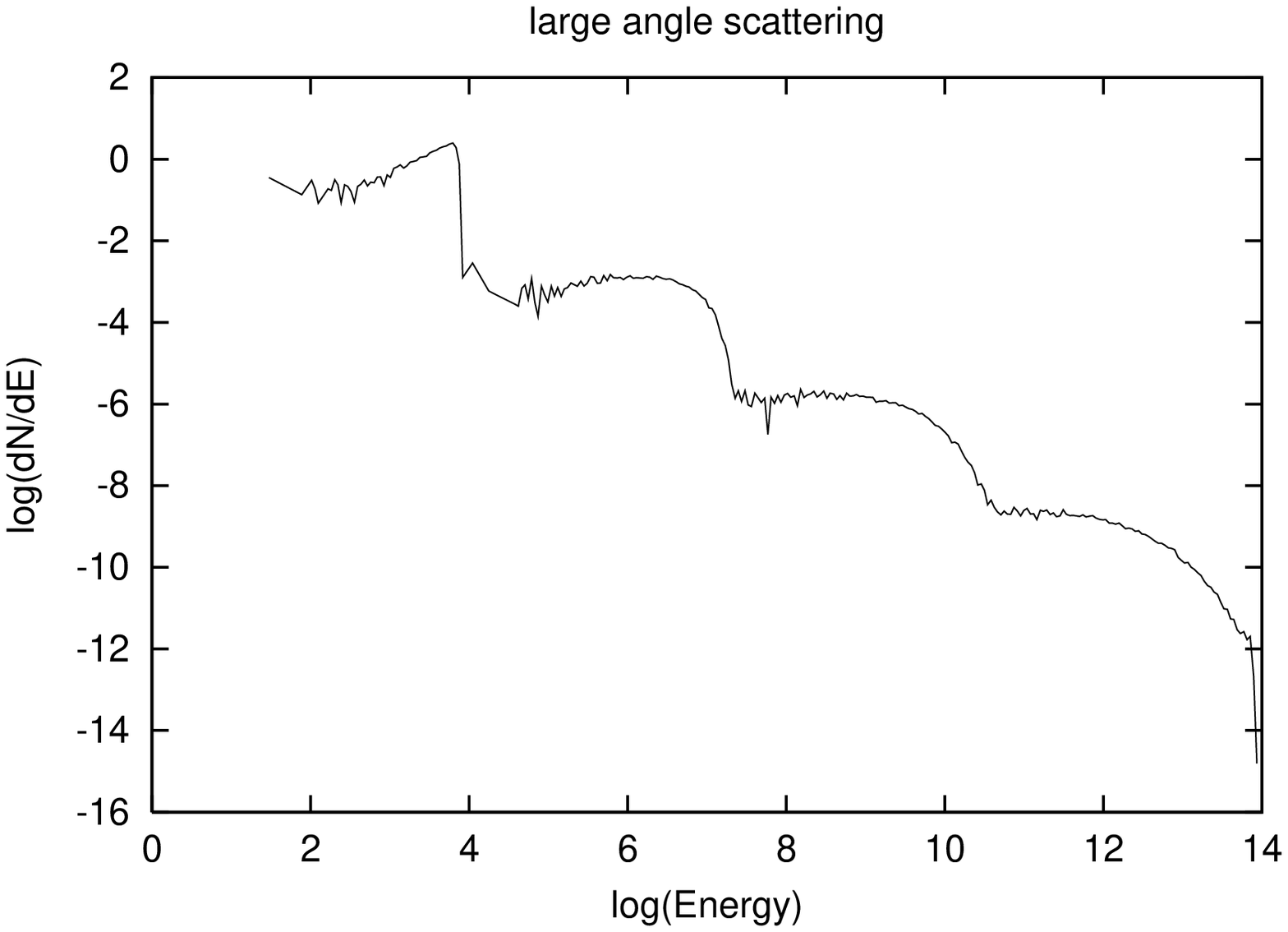,width=6.8cm}
\epsfig{figure=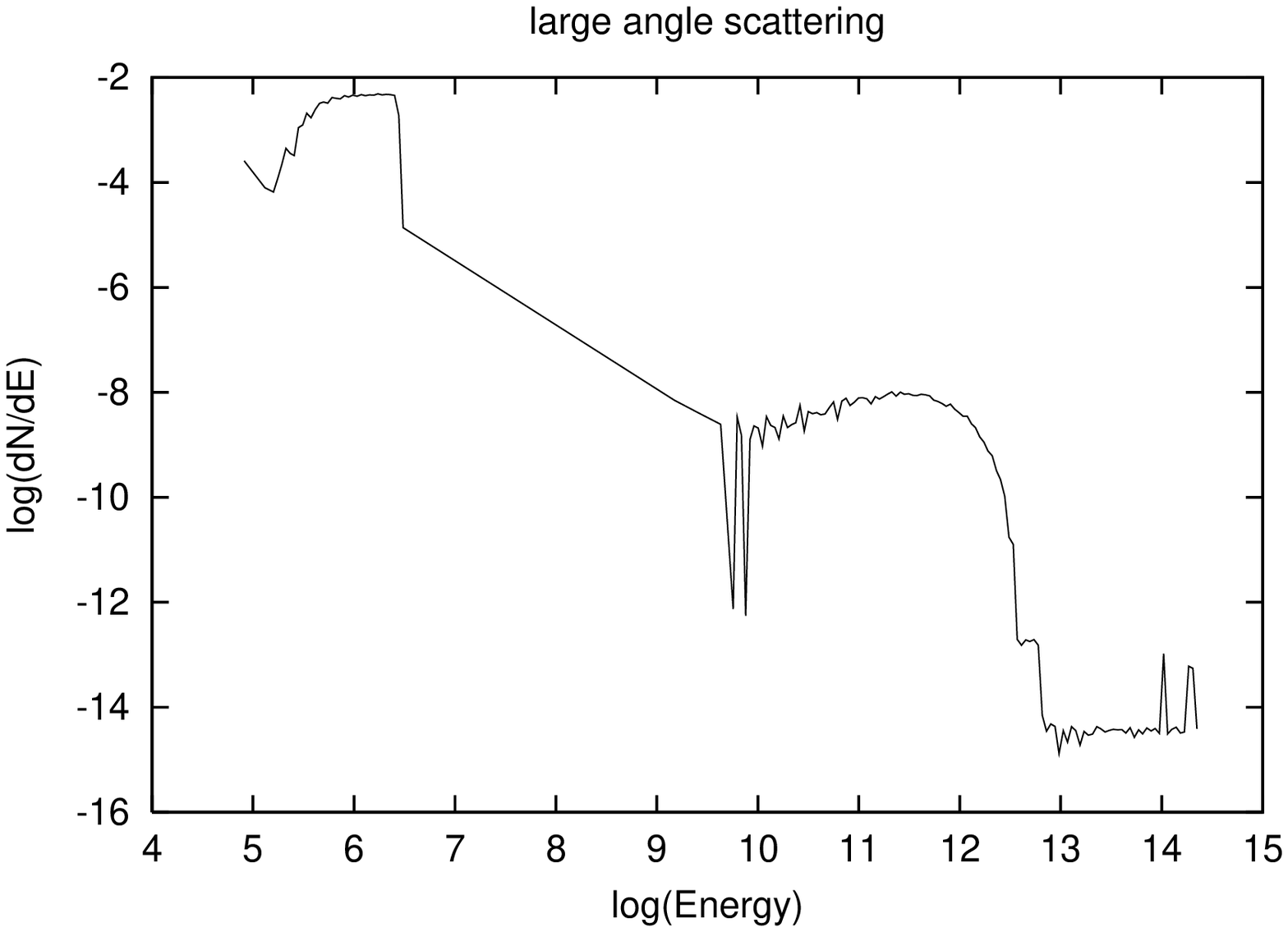,width=6.8cm}
\caption{Top plots: Spectral shapes for upstream gamma flow equal to 50 and large angle scattering.
$\psi$=15$^{\circ}$, $r$=4 (left), $r$=3 (right). Bottom plots: Gamma=50 (left), 990 (right), $r$=4, 
$\psi$=35$^{\circ}$.}
\end{center}
\end{figure}
The plateau-like shapes are again prominent and more 'disrupted', compared to parallel shock cases 
found in the similar computations of Meli and Quenby (2002a), for large angle scatter. 
In the case of pitch angle diffusion the shapes are much smoother,
particularly at lower $\Gamma_{1}$.  
The disrupted shapes are probably due to high anisotropy allowed in some situations.
We qualitatively understand the spectral shapes in terms of each plateau
exhibiting the acceleration gain in one shock cycle, down to up to down, with a $\Gamma_{1}^{2}$ factor 
boosting energy, followed by a high probability of loss downstream due to the beaming of the particles away 
from the shock.
Examples of the  beaming are presented in figure 9 where, we see the angular distributions of the transmitted
particles at the shock front in the de Hoffmann-Teller frame. We only show particles which have just
crossed the shock, not the complete, time averaged, distribution function.
The observed effect
is due to the anisotropies in pitch angle space expected for highly relativistic flows,
which are in accordance with lower $\Gamma_{1}$ results of Ellison et al.,
(1990) and Lieu et al., (1994).
\begin{figure}[h!]
\begin{center}
\epsfig{figure=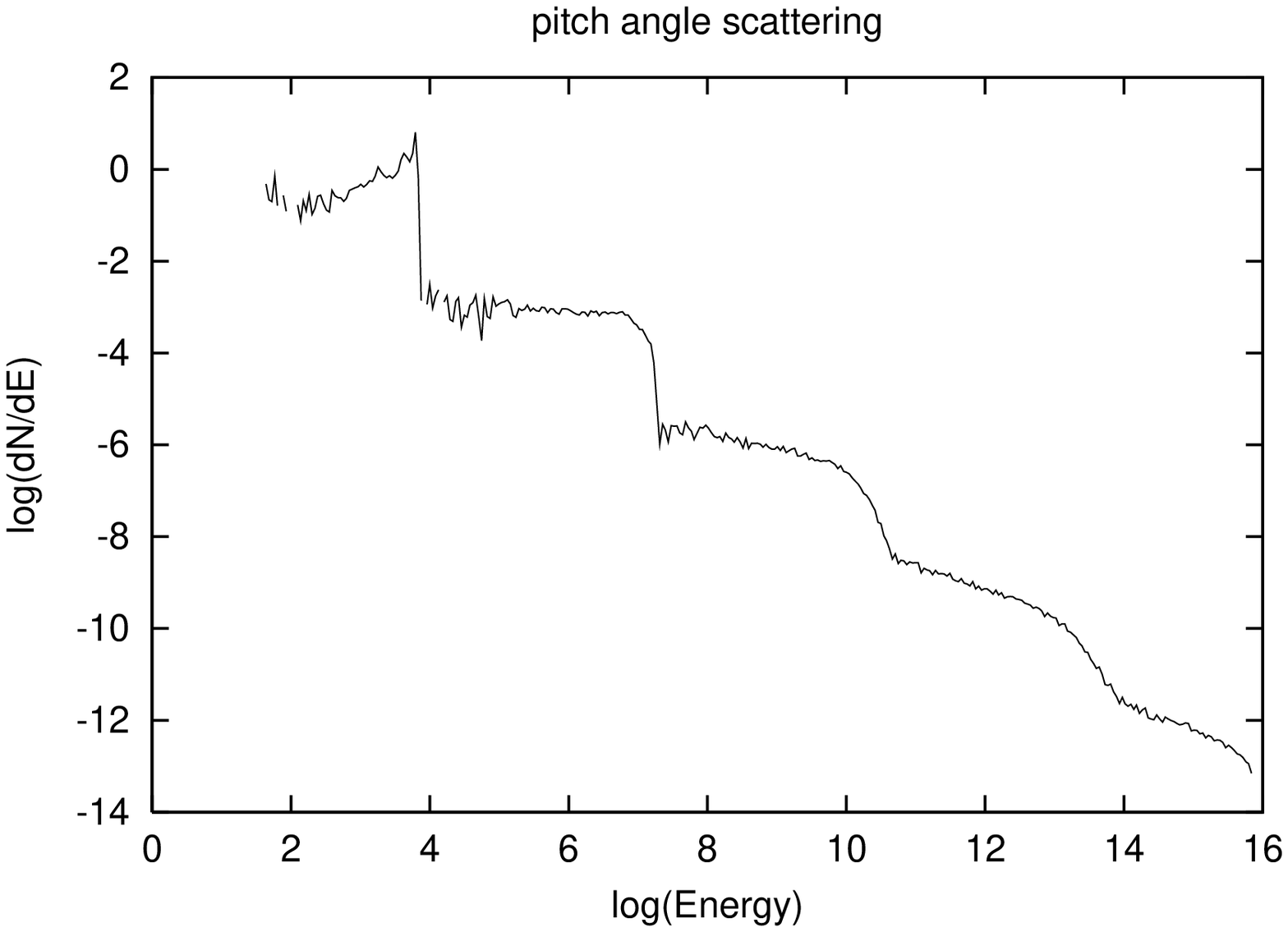,width=6.8cm}
\epsfig{figure=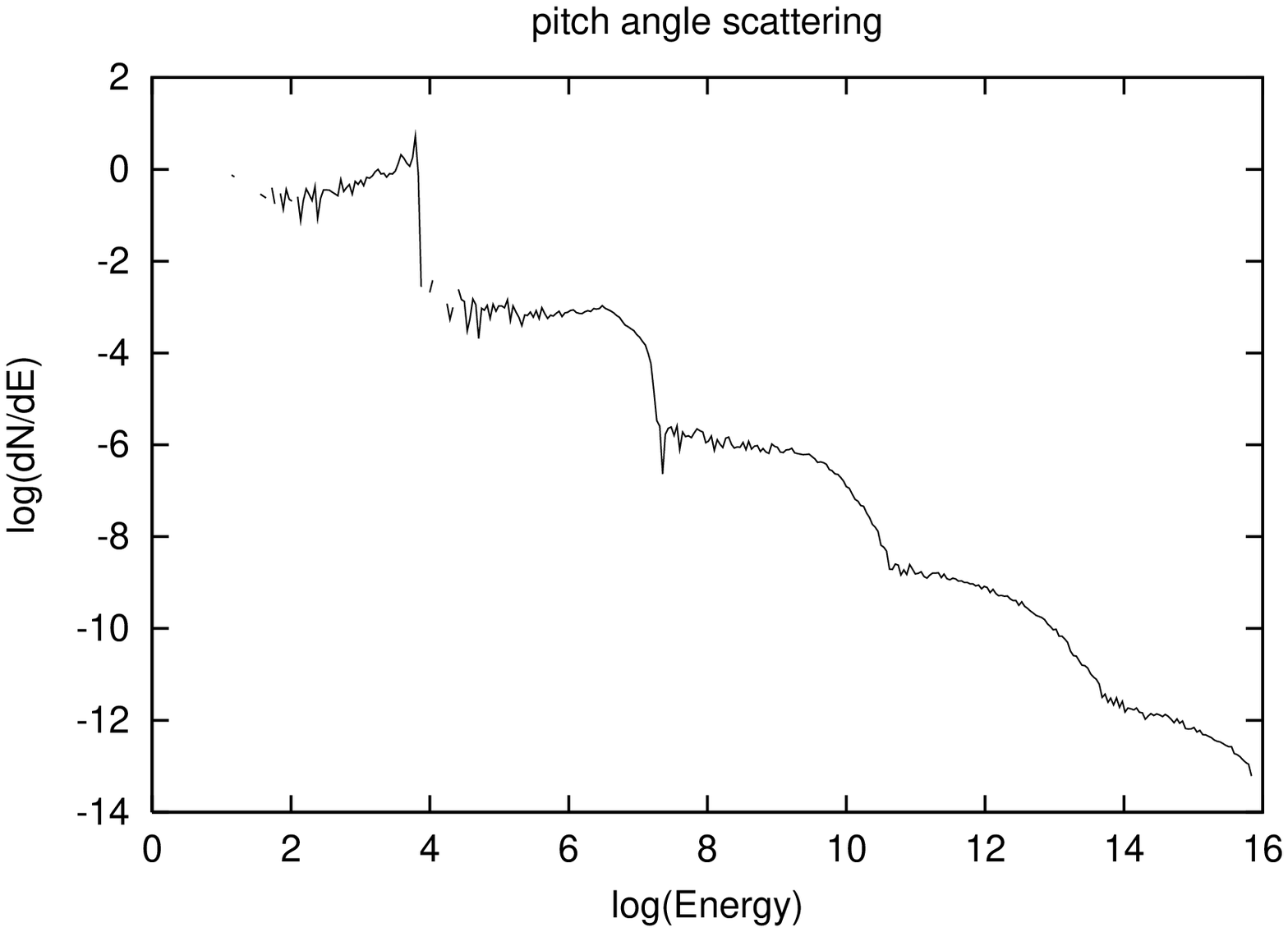,width=6.8cm}
\end{center}
\end{figure}

\begin{figure}[h!]
\begin{center}
\epsfig{figure=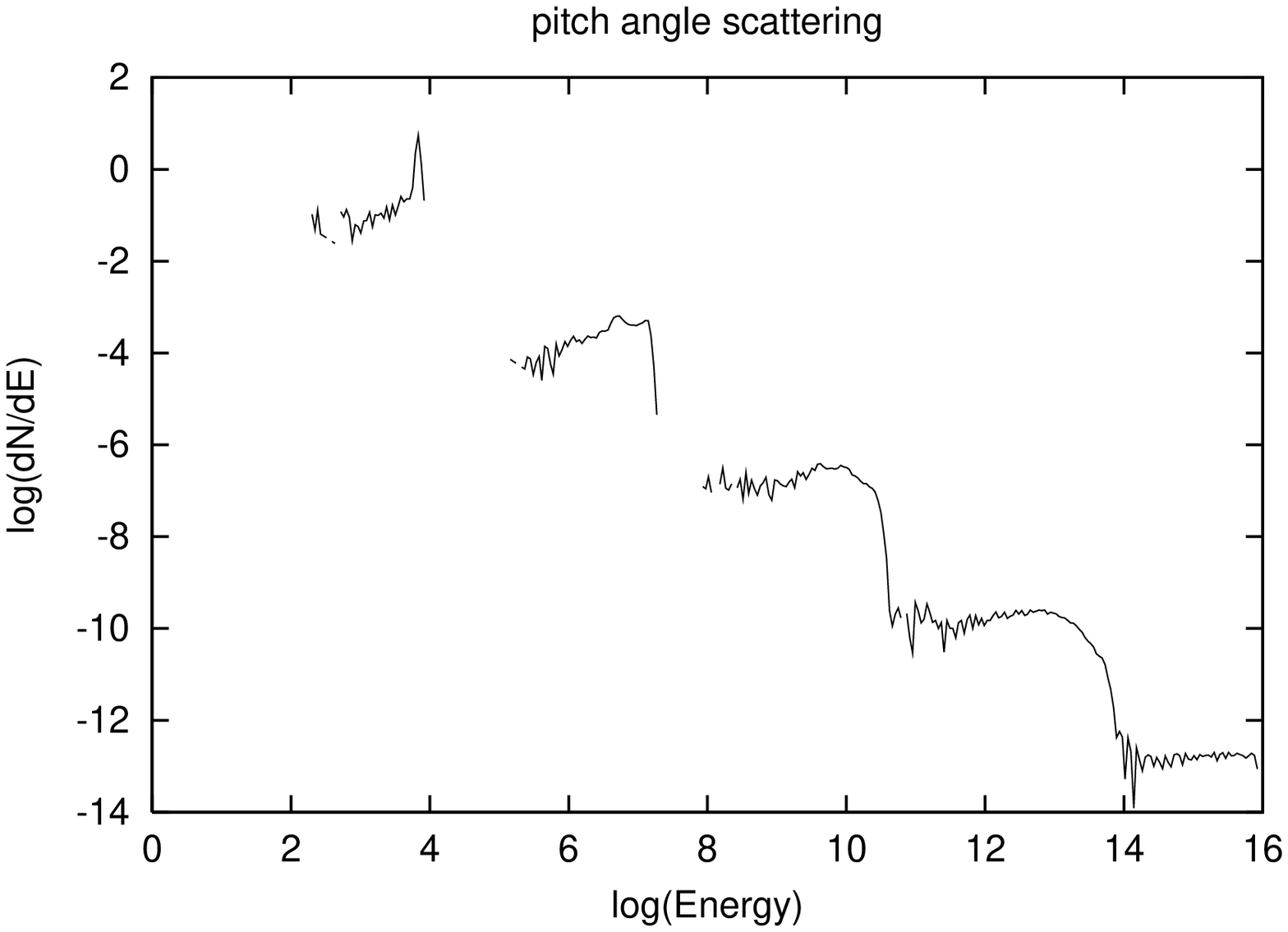,width=6.8cm}
\epsfig{figure=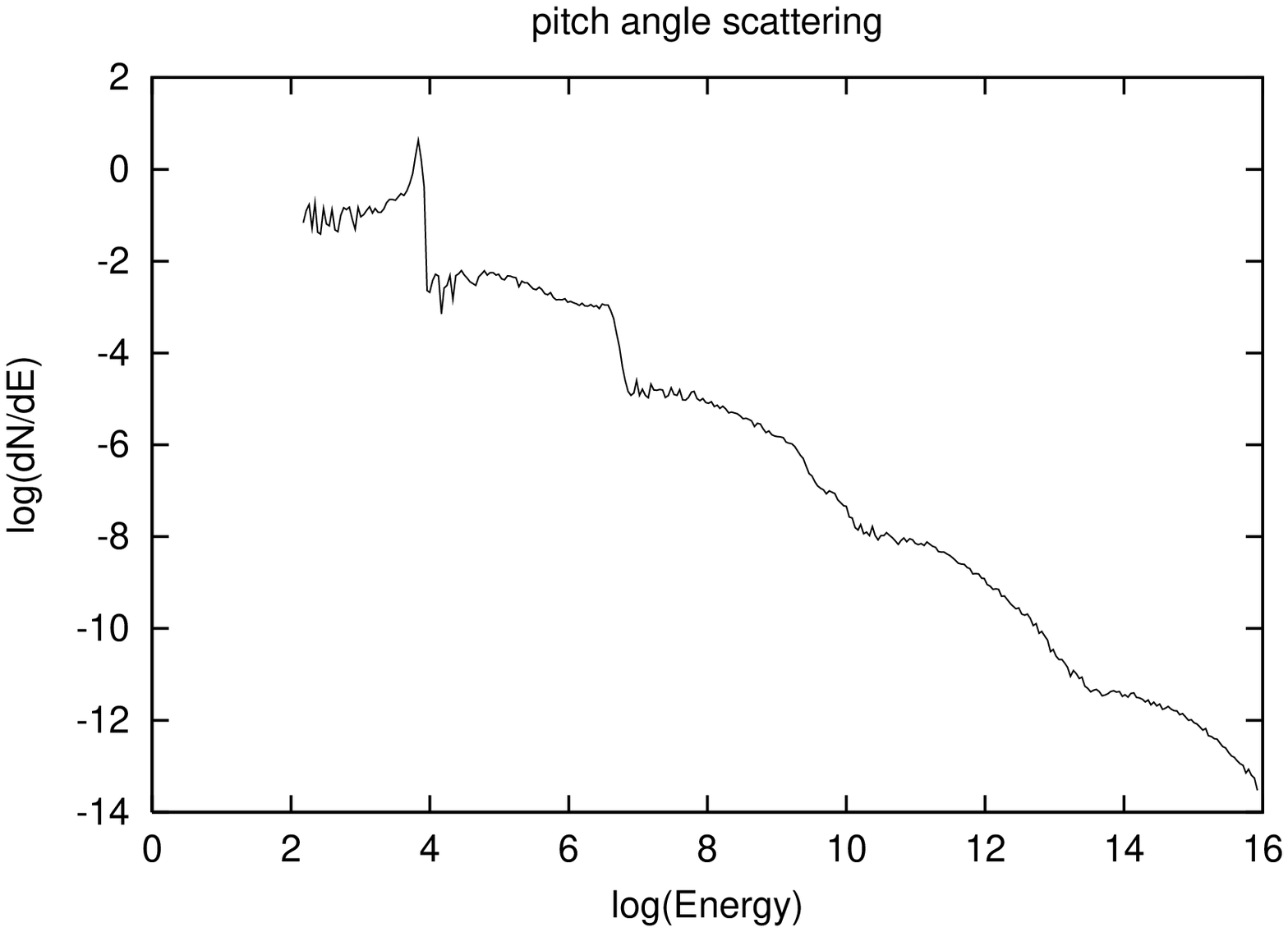,width=6.8cm}
\caption{Spectral shapes  for pitch angle diffusion, for upstream
$\Gamma$=50. Top plots: $\psi$=15$^{\circ}$, $r$=4 (left), $r$=3 (right).
Bottom ones: $\psi$=35$^{\circ}$, $r$=4 (left), $r$=3 (right).}
\end{center}
\end{figure}
These results show that, due to very relativistic velocities the distribution functions are very anisotropic 
and so the spatial diffusion approximation cannot apply.
In figures 10-11 we present results for the super-luminal case, where
$\psi_1= 48^{\circ}$ and $75^{\circ}$ and 
Lorentz factors range from 10 to 40, expecting to apply in the relativistic
flows of AGN environments. 
\begin{figure}[t!]
\begin{center}
\epsfig{figure=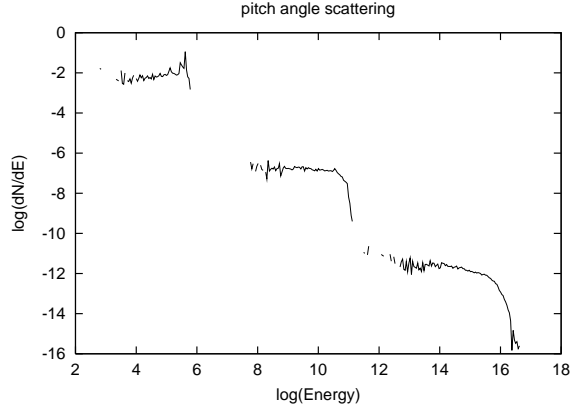,width=7.8cm}
\caption{Spectral shapes for $\Gamma=500$, $\psi$=15$^{\circ}$, $r$=4 and  pitch angle diffusion.}
\end{center}
\end{figure}
The compression ratio is equal to 4 and large angle scattering is used.
\begin{figure}[t!]
\begin{center}
\epsfig{figure=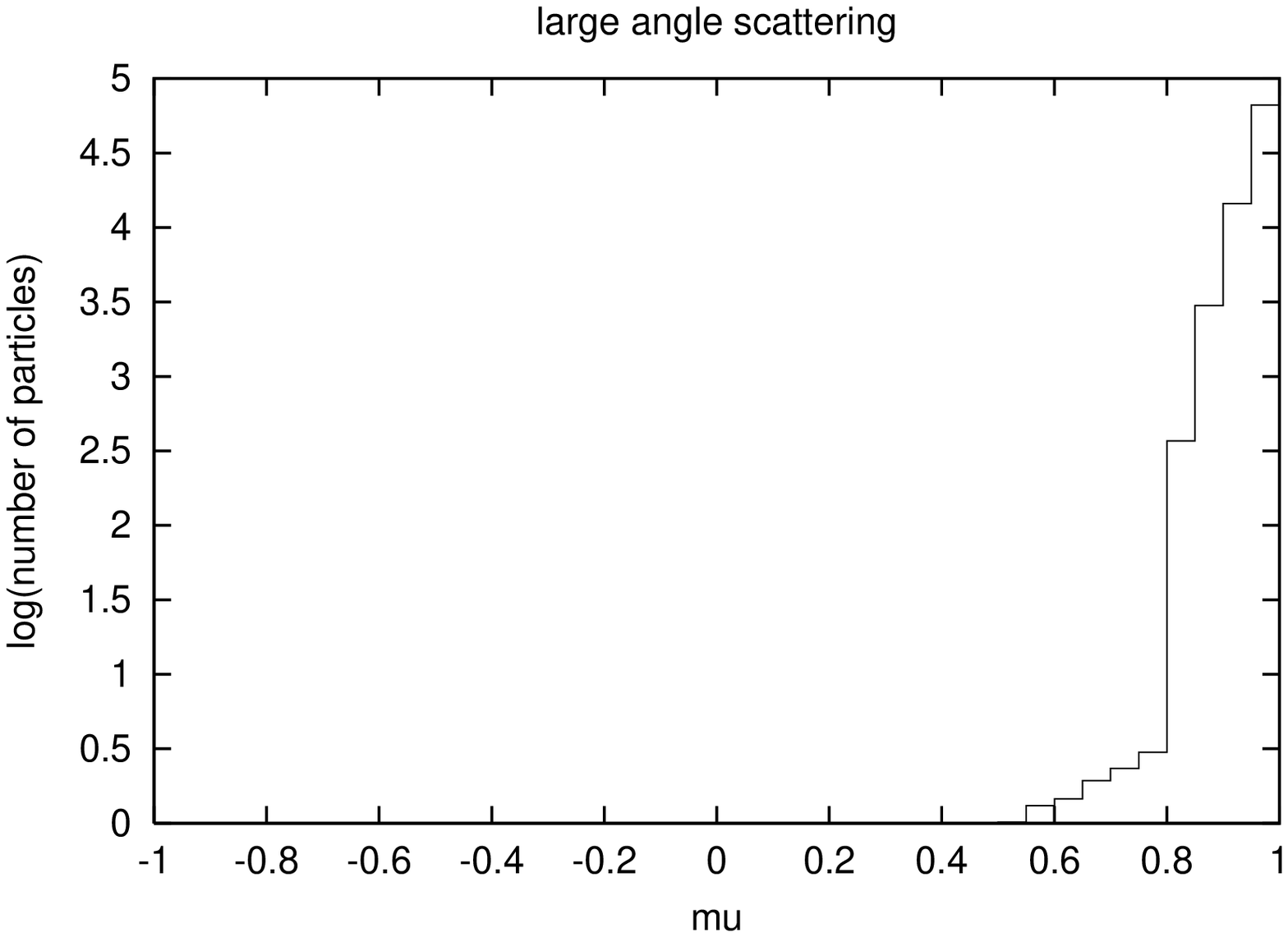,width=6.8cm}
\epsfig{figure=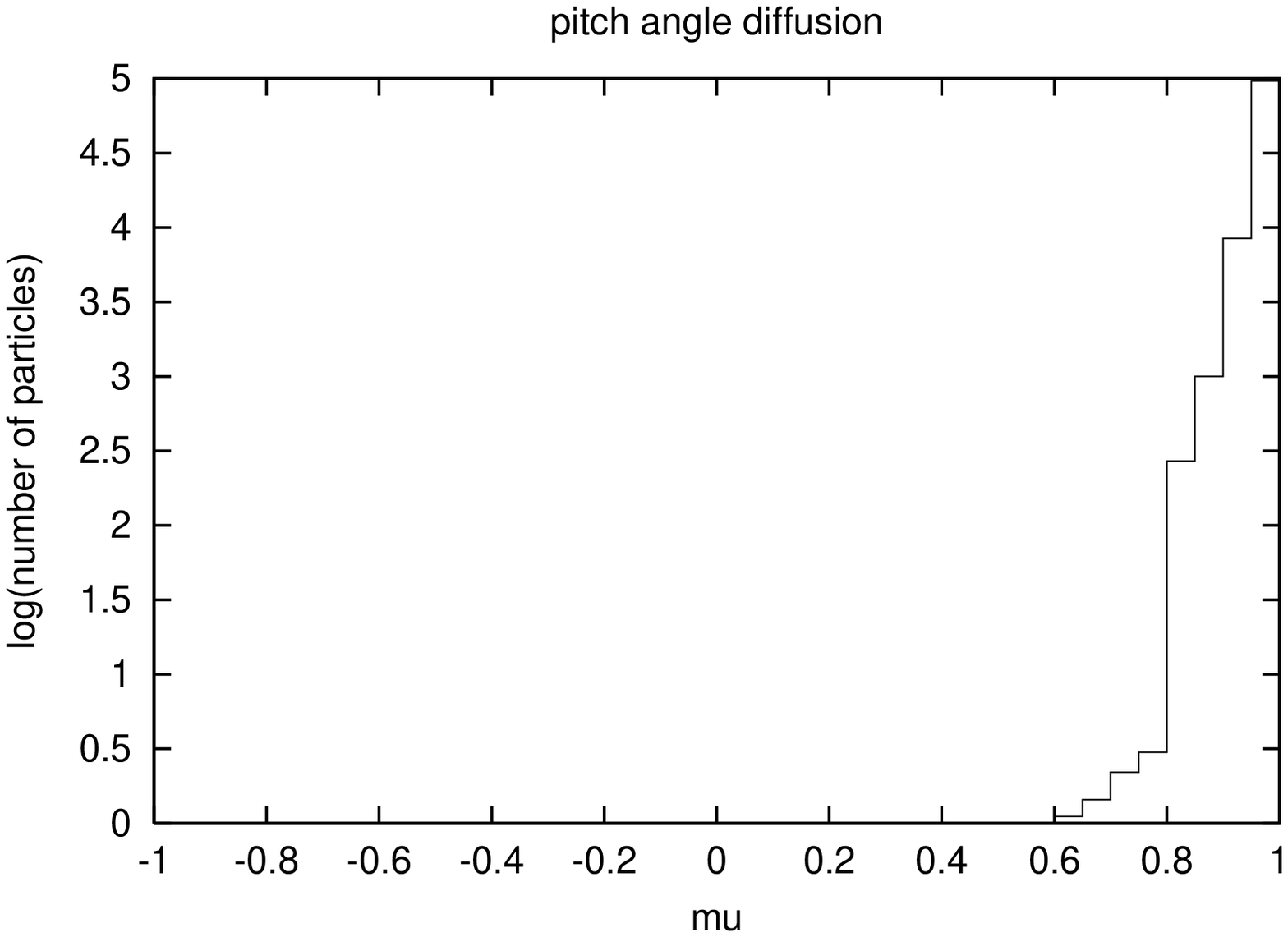,width=6.8cm}
\end{center}
\end{figure}
\begin{figure}[h!]
\begin{center}
\epsfig{figure=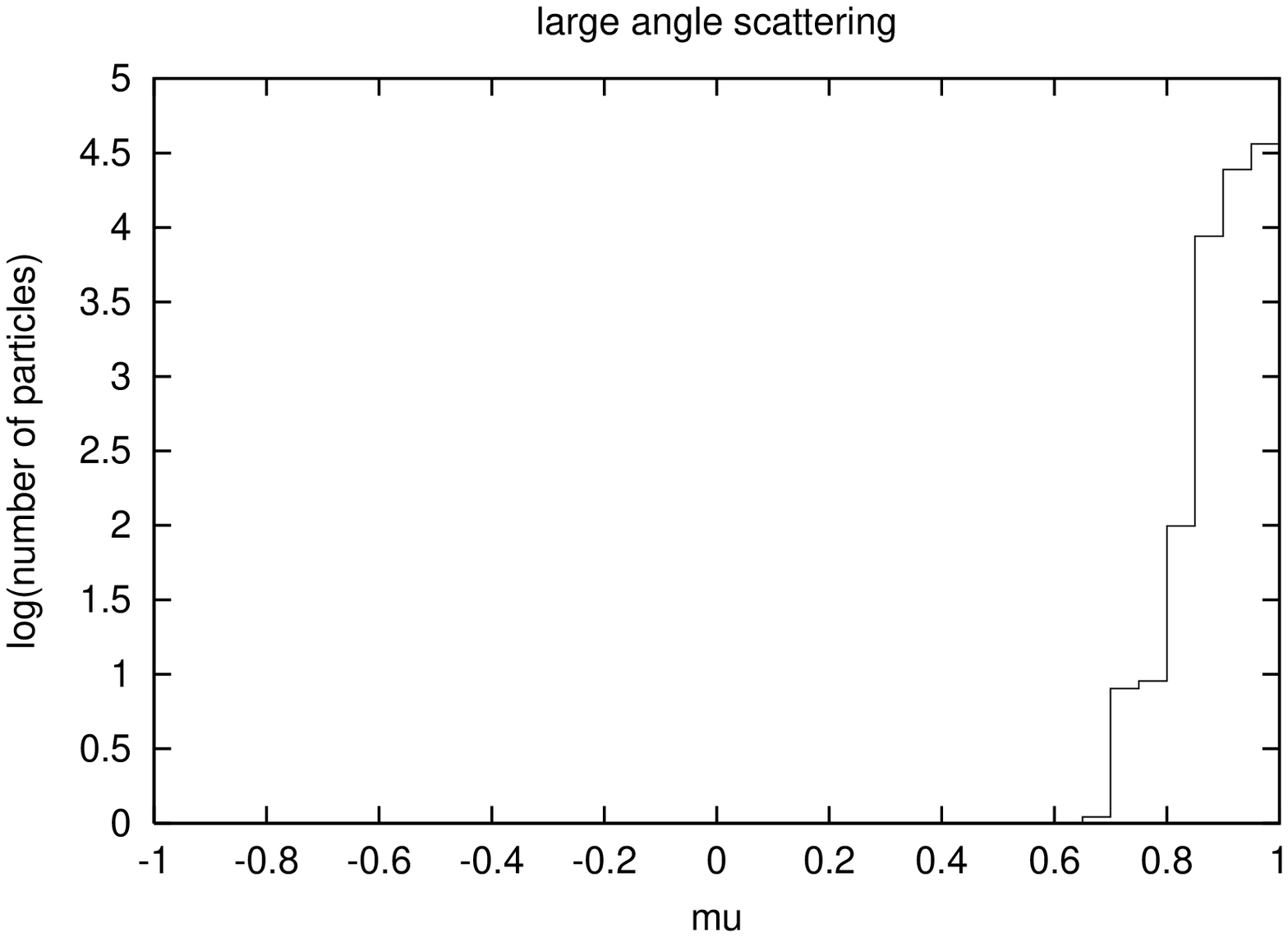,width=6.8cm}
\epsfig{figure=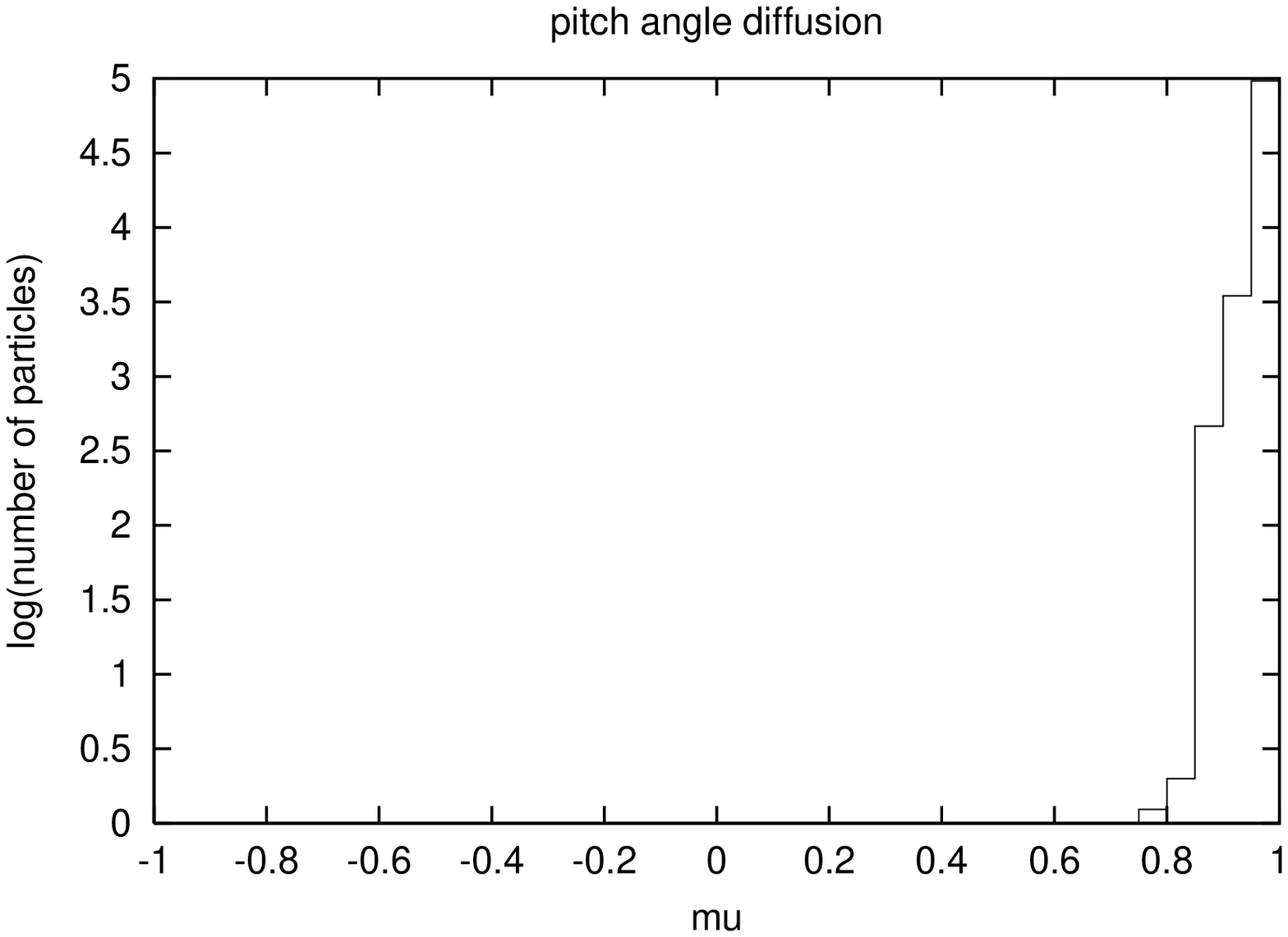,width=6.8cm}
\caption{The angular distribution of the logarithm of the number of transmitted (just crossed the shock)
particles versus mu = $\mu=cos\theta$ at  the de Hoffmann-Teller frame. Top plots: Upstream  flow $\Gamma$=200
$\psi$=15$^{\circ}$ and $r$=4. Bottom plots: Upstream flow $\Gamma$=990,  $\psi$=35$^{\circ}$ and $r$=4.  
In both cases we observe the strong 'beaming' of the particles' distribution.}
\end{center}
\end{figure}

\begin{figure}[t!]
\begin{center}
\epsfig{figure=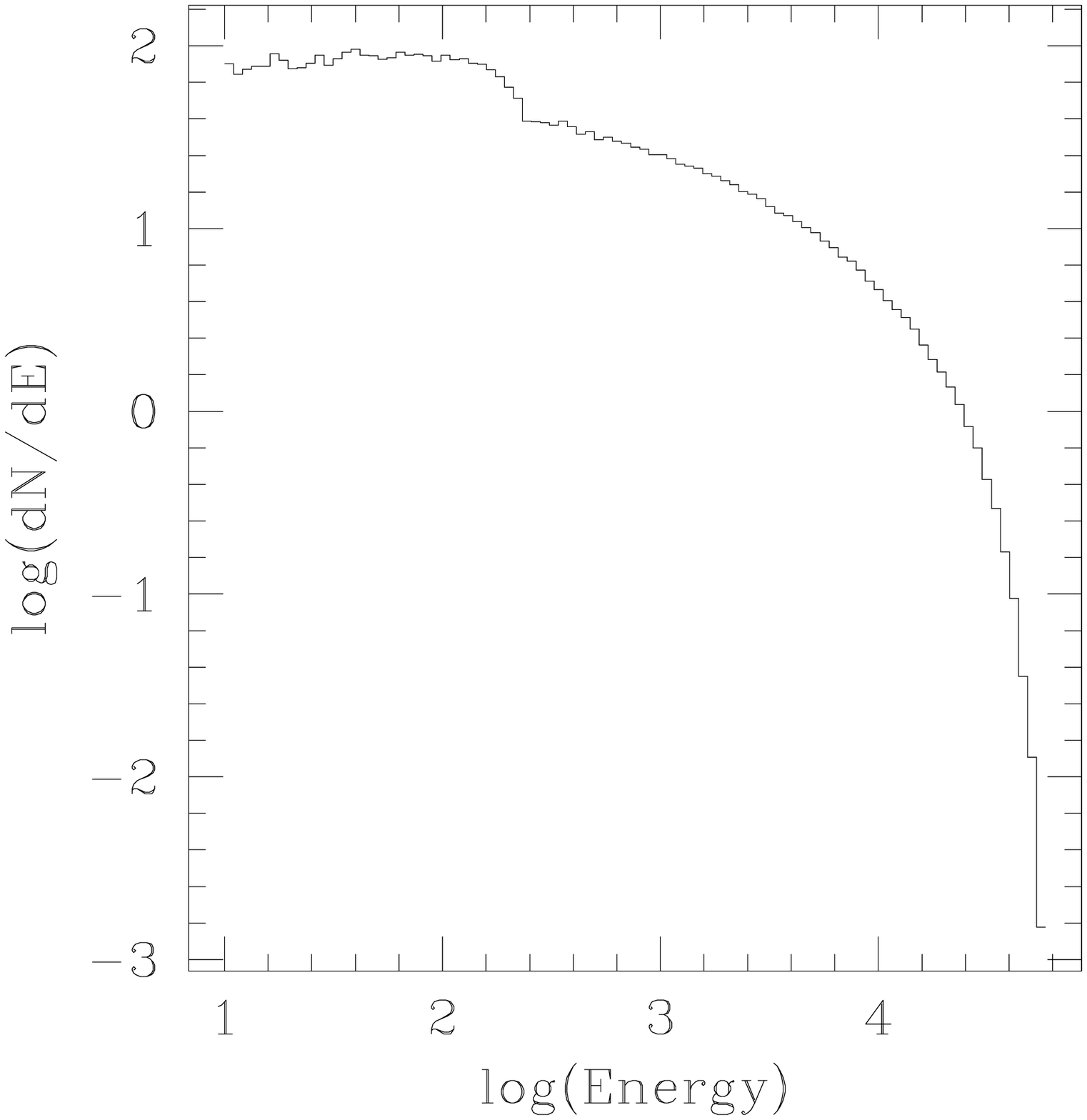,width=5.0cm}
\epsfig{figure=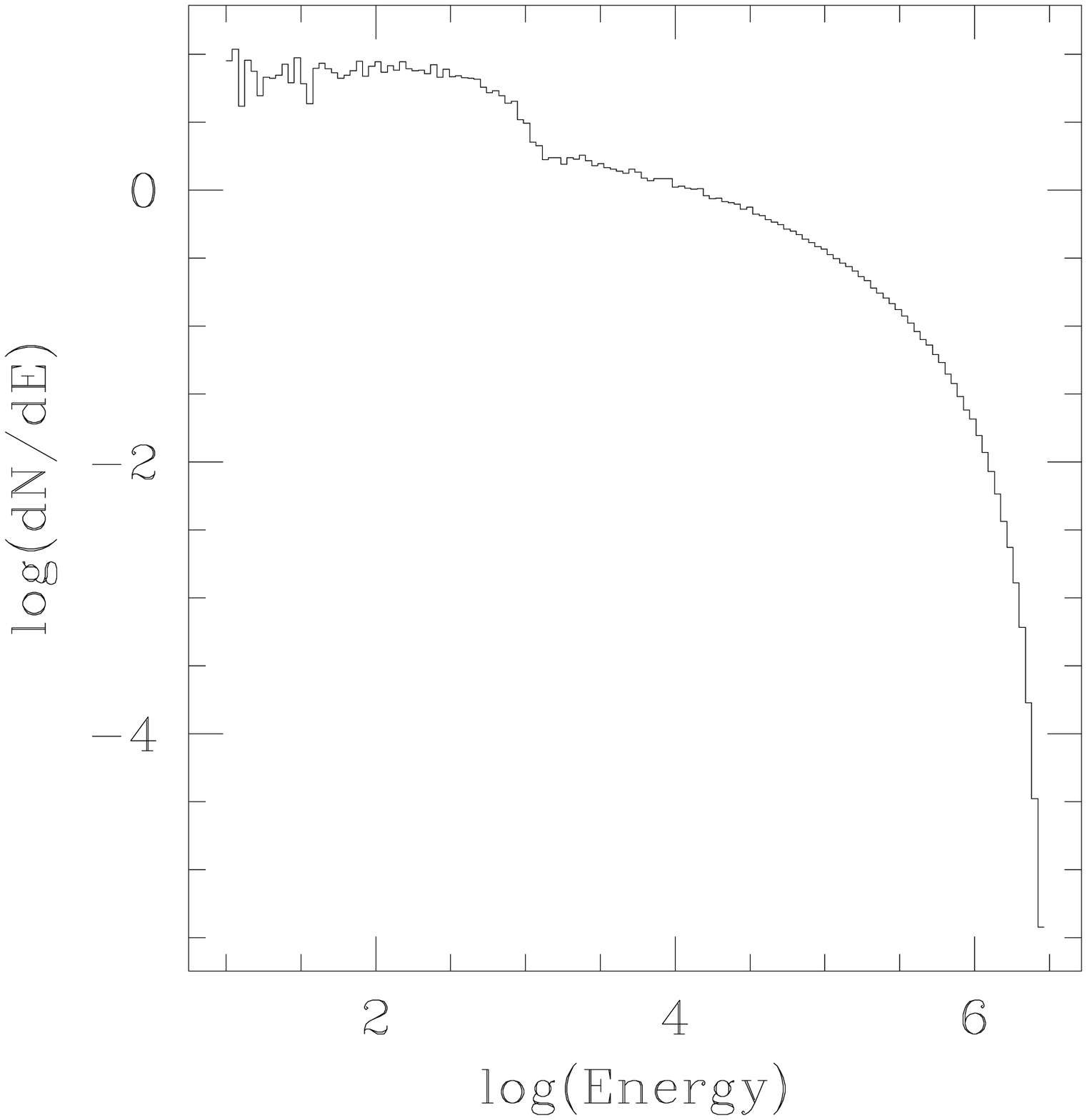,width=5.0cm}

\caption{Spectral shape for the super-luminal case, in 
the shock frame at the downstream side for $\Gamma$=10 and $\psi$=48$^{\circ}$ (left), $\Gamma$=30 and 
$\psi$=75$^{\circ}$ (right).}
\end{center}
\end{figure}
To obtain the spectral shapes for the accelerated particles
as the trajectories are followed during the simulations, we
add the momentum of each particle in the corresponding bin every time
the particle crosses the shock as before to obtain the mean energy gain and distribution function.
We observe that the spectral shape falls away from a
power law and descends rapidly to an upper cut off which is expected, because the
particles are swept away by the flow after one cycle and have a very limited chance of return upstream,
especially due to the high field inclination to the shock normal.
\begin{figure}[t!]
\begin{center}
\epsfig{figure=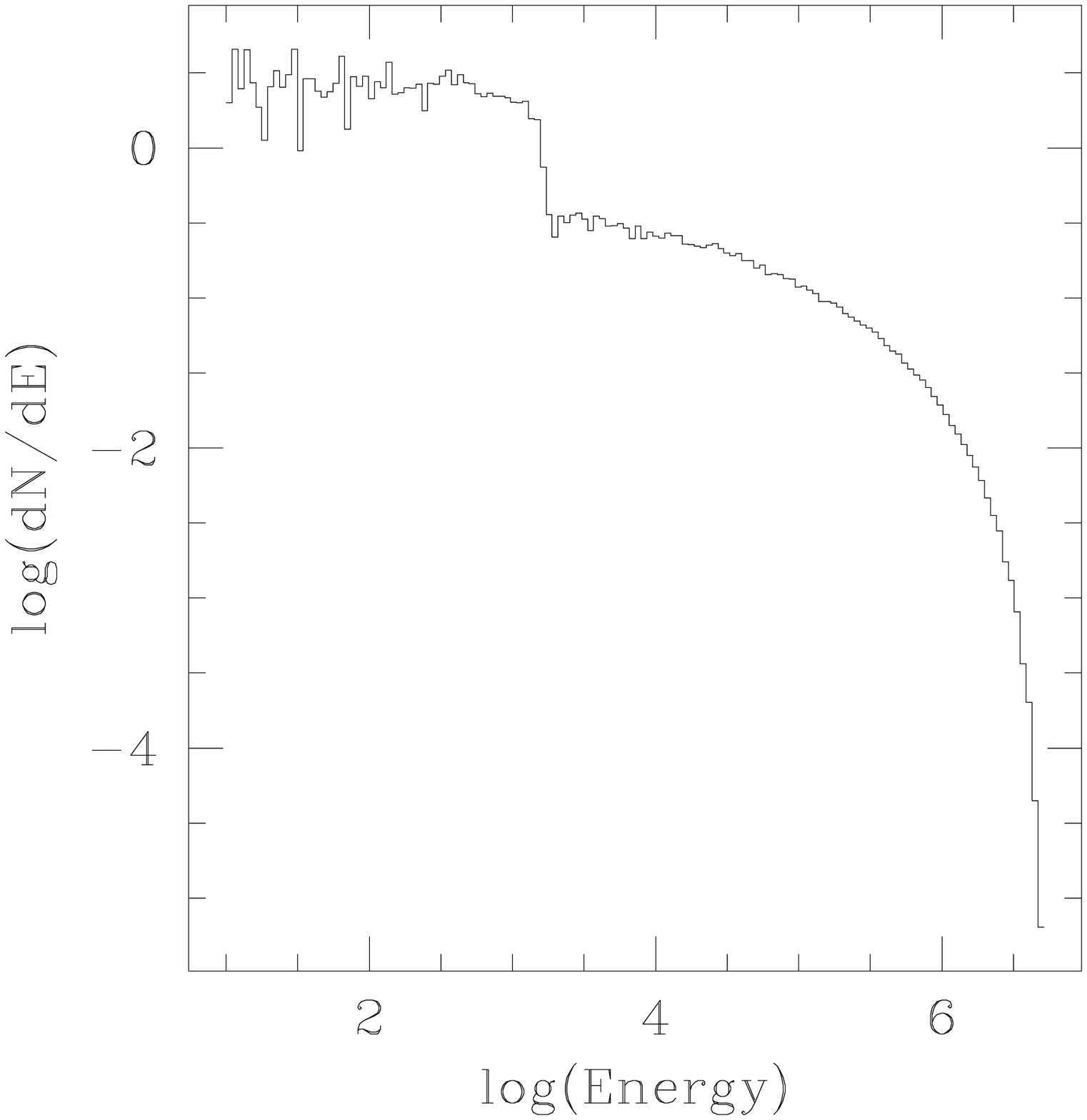,width=7.0cm}
\end{center}
\end{figure}

\begin{figure}[h!]
\begin{center}
\epsfig{figure=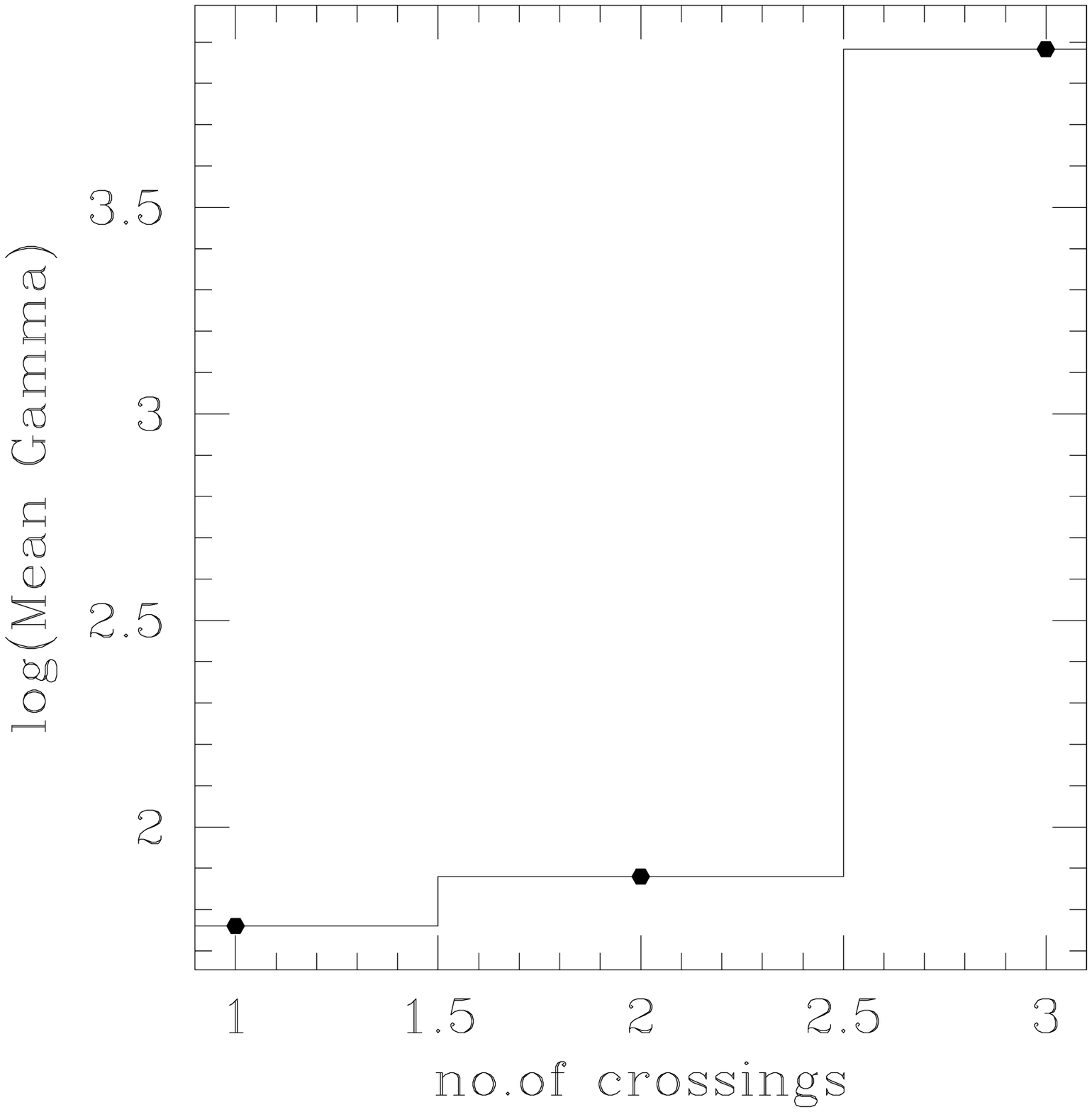,width=5.0cm}
\epsfig{figure=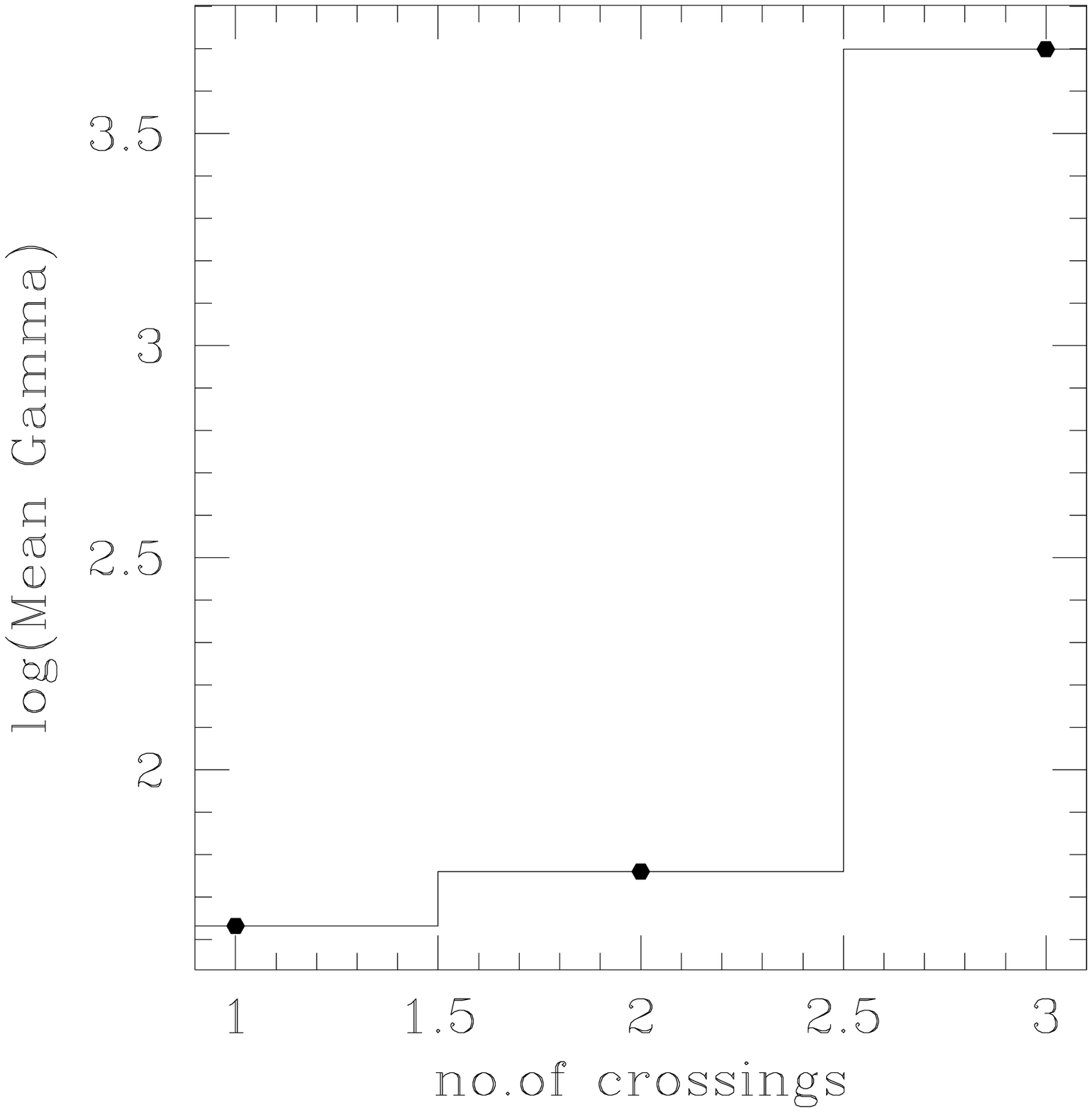,width=5.0cm}
\caption{Top plot: Spectral shape for the super-luminal case in the shock frame at the downstream side 
for $\Gamma$=40 and $\psi=75^{\circ}$. 
We observe a structure at the  beginning of the spectrum -near the injection point- and a cut off at 
high energies, compared to typical diffusive shock acceleration spectra.
Bottom plots: Energy gain versus number of shock crossings (1-2-3) of the particle's 
helix trajectory  for $\Gamma$=10, $\psi$=48$^{\circ}$ (left) and $\psi$=75$^{\circ}$ (right).}
\end{center}
\end{figure}
In the lower plots of figure 11, the energy gain of the particle is shown, versus the number 
of the helix-trajectory shock crossings for Lorentz factor 10 and 
$\psi_1= 48^{\circ}, 75^{\circ}$  between the magnetic field and the shock normal.
It is seen that after the 3$^{rd}$ crossing (after one cycle) the particle gains
a considerable amount of energy ($\sim \Gamma^{2}$), but further gain is limited due to the fact
that the particles are swept away by the flow very rapidly with no chance of
're-cycling'  the shock again. Our calculations show that $\sim60\%$ of the particles 
cross the shock three times (upstream$\rightarrow$downstream $\rightarrow$ upstream 
$\rightarrow$downstream).
It is possible to conclude that 'one-shot' drift acceleration has taken place,
that is on one up to down crossing and there is energy increase as
expected from drift in the shock frame electric
field under the magnetic field gradient. Some particle return upstream has been possible 
under the action of large angle, downstream scattering.
However this return flux is so limited that a power law spectrum cannot 
develop and a cutoff is soon reached. 
Computational time limitations have currently prevented an extension of the modeling to 
higher $\Gamma$ flows and pitch angle scattering models for the super-luminal case.


\section{Discussion and Conclusions}

Oblique particle shock  acceleration simulations for sub-luminal and super-luminal
cases have been presented. High gamma upstream flows have been used ranging from 
$\Gamma$=10 up to $\sim 10^{3}$.  The results found are important in answering 
questions concerning the efficiency and relativistic effects of first order Fermi acceleration. 
A major finding of both this work and the companion paper (Meli and Quenby, 2002a),
is that in contrast to parallel shocks with small angle scattering, all other
types of shocks and scattering models do not yield smooth, power law spectra 
at the shock in the $\Gamma>>1$ regime, even though subsequent smoothing 
during particle escape may arise. Reasons for this will be re-iterated later.
However, sub-luminal, inclined relativistic shocks provide spectra more extended in energy 
and therefore more approximating to a power law distribution than super-luminal shocks, the latter 
exhibiting a pronounced high energy cutoff feature. The result that up to 3 crossings of the shock can 
often occur in the adopted large angle scattering model 
implies that the super-luminal case is not simply 'one-shot'
shock drift. However the large gain crossing $2 \rightarrow 3$ followed by loss to the system can well 
be described as the energy gain experienced on the last shock drift pass through the discontinuity 
standing out from the smaller, average diffusive gain.\\
Concerning the important 'speed up' effect in the sub-luminal case,
this appears to be due to a combination of $\Gamma^{2}$ or $\Gamma$ energy gain 
and the anisotropic angular distributions at the shock front 
(Ellison et al., 1990) which as we showed, is confined to a very small angle 
transmission cone. Thus, only a narrow range of pitch angle particles return to the shock 
and are 'counted' for the acceleration time calculation, as the ones with larger pitch angles
are lost downstream. If the particle distribution is made isotropic each side of the shock before re-crossing,
the energy gain per cycle is $\sim\Gamma_{1}^{2}$ Quenby and Lieu (1989), but a distribution remaining 
anisotropic
will experience less energy gain (Gallant and Achterberg, 1999). Hence some 
part of the 'speed up' compared with the non-relativistic situation is expected to be connected with 
the un-allowed for energy enhancement.
The anisotropy is of course linked to the 'beaming effect' found also in 
mild relativistic cases, in accordance with Peacock (1981) and
Kirk and Schneider (1988). Lieu and Quenby (1990) and Lieu et al. (1994) note that 
the distribution of the particle's pitch angle is peaked 
towards the direction of the  relativistic transformation to the de Hoffmann-Teller 
frame. In other words this beaming which enhances the number of particles close to the
field  line, may provide an additional reason for the decrease in acceleration time as the flow becomes more 
relativistic for both large angle and pitch angle diffusion.
This 'speed-up' resembles a trend found in the parallel shock particle acceleration case. 
Furthermore, the last arguments could explain the 
prominent spectral 'plateau' shapes found in many oblique, sub-luminal and parallel shock 
cases where each shock cycle crossing seems to stand  out \textit{separately}. Enhanced single cycle 
acceleration is followed by large flux loss downstream, preventing statistical smoothing into a single power 
law accelerated spectrum.\\
The 'speed up' effect -which has also been observed in the mildly relativistic 
regime Quenby and Lieu (1989) and Ellison et al. (1990), is important for a 
complete understanding of particle acceleration to the highest energies. An understanding of 
this crucial effect is necessary in any explanation for the maximum energy that particles 
(protons, electrons, iron nuclei) can  achieve in GRB fireballs, AGN hotspots, and pulsar 
ultra relativistic polar winds. 
For example results found  in the previous sections dealing with the sub-luminal case for $\Gamma\sim 10^{3}$, 
show only few crossings are needed for 
the particle to achieve energies up to $10^{19}-10^{20}$eV, although for $\Gamma\sim$ 10 more than ten crossings
may be needed to gain such high energies. It is the competition between loss time and energy gain time 
which determines the maximum energy in many situations. Generally speaking the percentage increase of the
maximum energy would depend on many factors. One of those is the velocity of the upstream flow. 
Also the dimensions of the shock could play a vital role in order to define the 
escape losses in a specific shock acceleration site. In addition another
factor in the physical conditions which could alter the cosmic ray
spectrum and decrease the spectral index is due to different energy losses occurring in
the regime; for example the presence of high or low energy photons (Greisen-Zatsepin-Kuz'min cutoff 
effect{\footnote {This is called also the microwave background effect, where the interaction between 
the cosmic rays and the microwave background results 
in a cutoff in the energy spectrum of cosmic rays (Greisen 1966; Zatsepin-Kuz'min 1966)}}) 
or $\gamma-\gamma$ flux interactions{\footnote{If GRB high energy photons come from electron acceleration in the
$\Gamma\sim 10^{3}$ regime, we might expect some 'structure' in the spectrum at the higher 
energies on any scattering model.} in GRB will decrease
the maximum energy of the particles.
In the context of AGN it is well known from a number of interesting papers such as 
Protheroe and Kazanas  (1983) and Chakrabarti and Moltoni (1993) that a shock in the
central engine of the accreting flow ($\Gamma \simeq 5-10$) is likely to appear
and particle acceleration can consequently take place. Also there is observational evidence 
that a common feature of many AGN is the formation of one or two jets where shock formations 
should appear in conjunction with $\Gamma\sim 5-10$ flows. Thus, the computed flattening of the 
spectra -they are all flatter than a -2 power law-  as well as the acceleration 'speed up', should have important 
consequences when compared with predictions and observational data, regarding for example,
the photons produced by emission from electrons which  can be accelerated at the shocks or produced from 
accelerated protons.
GRB seem to be another potential candidate for the acceleration of Ultra High Energy Cosmic Rays (UHECRs), 
but the issue is debated by many workers.  As we have mentioned Vietri (1995) and 
Waxman (1995), who have exploited the 'gamma-squared' factor energy gain of the 
particle, were first to predict theoretically that indeed UHECRs can be produced 
in GRB where a pre-acceleration of the particles has been implied.
Briefly, we note that observationally there is very strong evidence that 
pulsars (e.g. PSR 1913+16, PSR 2127+11C) are capable of injecting continuously
relativistic particles into the  surrounding medium over their lifetime. While this 
plasma will probably be predominantly in the form of $e^{+}e^{-}$ pairs created in the pulsar 
magnetosphere, it has been argued that pulsar winds must also contain ions in order to account 
for the electrical current required in the Crab Nebula (Hoshino et al., 1992; Gallant and Arons, 1994). 
The termination shock as the $\Gamma\sim 10^{5}$ flow meets the nebula is likely to be
quasi-perpendicular and hence the super-luminal situation applies for the additional
$e^{+}e^{-}$ acceleration.
The enhanced production of neutrinos in astrophysical sites such as in AGN and 
GRB is another consequence of the simulation findings of the 'speed up' effect and 'gamma 
squared' factor.  Mastichiadis and Kirk (1992),  Protheroe and  Szabo (1992) and Vietri (1998) 
showed that the neutrino flux depends on the maximum energy attained from the primary protons, 
thus a decrease of the acceleration time constant could produce higher maximum energy limits for the
accelerated particles and consequently an increase in the cutoff energy of the neutrino flux will be noted.\\

{\noindent \bf Acknowledgements}\\

We would like to express our thanks to Prof. Drury and the unknown referee for valuable comments
and suggestions.


\end{document}